\documentclass[showpacs,aps,twocolumn]{revtex4}
\usepackage{epsfig}
\usepackage{graphicx}
\usepackage{amsmath,amssymb,amsfonts}
\usepackage{array}
\usepackage{url}
\usepackage{multirow}
\usepackage{float}
\usepackage{lineno}
\usepackage{xspace}
\usepackage{ulem}
\usepackage{bm}

\newcommand{\bef}{\begin{figure}}
\newcommand{\eef}{\end{figure}}
\newcommand{\bc}{\begin{center}}
\newcommand{\ec}{\end{center}}
\newcommand{\nn}{\nonumber}
\newcommand{\be}{\begin{equation}}
\newcommand{\ee}{\end{equation}}
\newcommand{\bea}{\begin{eqnarray}}
\newcommand{\eea}{\end{eqnarray}}

\def\ba{\begin{eqnarray}}
\def\ea{\end{eqnarray}}

\usepackage[usenames,dvipsnames]{color}
\definecolor{darkblue}{RGB}{0,0,196}
\usepackage[colorlinks=true,linkcolor=darkblue,citecolor=darkblue,urlcolor=darkblue]{hyperref}

\begin{document}
\title{Can charm fluctuation be a better probe to study QCD critical point?}

\author{Kangkan Goswami}
\email[]{Kangkan.Goswami@cern.ch}
\author{Kshitish Kumar Pradhan}
\email[]{Kshitish.Kumar.Pradhan@cern.ch}
\author{Dushmanta Sahu}
\email[]{Dushmanta.Sahu@cern.ch}
\author{Jayanta Dey}
\email[]{jayantad@iitbhilai.ac.in}
\author{Raghunath Sahoo}
\email[Corresponding Author: ]{Raghunath.Sahoo@cern.ch}
\affiliation{Department of Physics, Indian Institute of Technology Indore, Simrol, Indore 453552, India}

\begin{abstract}
    We study the diffusion properties of an interacting hadron gas and evaluate the diffusion coefficient matrix for the baryon, strange, electric, and charm quantum numbers. For the first time, this study sheds light on the charm current and estimates the diffusion matrix coefficient for the charmed states by treating them as a part of the quasi-thermalized medium. We explore the diffusion matrix coefficient as a function of temperature and center-of-mass energy. A van der Waals-like interaction is assumed between the hadrons, including attractive and repulsive interactions. The calculation of diffusion coefficients is based on relaxation time approximation to the Boltzmann transport equation. A good agreement with available model calculations is observed in the hadronic limit. To conclude the study, we discuss, with a detailed explanation, that charm fluctuation is expected to be a better tool for probing the QCD critical point.  

\end{abstract}
\date{\today}
\maketitle
\section{Introduction}

The primary goals of the Large Hadron Collider (LHC) at CERN and the Relativistic Heavy Ion Collider (RHIC) at BNL are to reproduce extreme conditions of the early stages of our universe in the laboratory. At very high energy densities and/or temperatures, the hadrons melt to produce a new state of deconfined matter called quark-gluon plasma (QGP). This new phase of matter is locally thermalized and exhibits collective behavior but exists for a very short period, around a few $\text{fm}/c$. In ultra-relativistic heavy-ion collisions at the LHC energies, QGP is created at substantially high temperatures ($T$) and almost zero or low baryon chemical potential ($\mu_{\rm B}$). In this regime, lattice Quantum Chromodynamics (lQCD) estimates the transition from the QGP to the hadronic matter as a crossover at $T \simeq 140 - 190$ MeV~\cite{Aoki:2006we}. Conversely, a first-order phase transition might be expected at very high $\mu_{\rm B}$ and low $T$. By reducing the collision energy ($\sqrt{s_{\rm NN}}$) of the colliding beams, one can explore the large $\mu_{\rm B}$ region, corresponding to the RHIC, FAIR, and NICA experiments.  There also exists a possible critical point, where both the crossover and the first-order lines meet~\cite{Stephanov:1998dy}. Such conditions are investigated with the RHIC Beam Energy Scan (BES) program~\cite{STAR:2010mib,STAR:2013gus,STAR:2014egu}, which is a fascinating field of research nowadays. To locate this QCD critical point, one can study the fluctuations of conserved charges: baryon ($B$), strangeness ($S)$, and electric ($Q$) charges. They play a vital role in finding the critical point around which fluctuations of conserved charges can exhibit non-monotonic behavior~\cite{Asakawa:2000wh,Jeon:2000wg,HotQCD:2012fhj,HotQCD:2017qwq}. The theoretical description of QCD matter at finite $\mu_{\rm B}$, especially near the critical point, requires an appropriate description of the transport of conserved charges.  Moreover, due to the rapid expansion of the fireball, fluctuations originating in the QGP phase may survive until freeze-out, thus serving as a signal of QGP formation in the early stages of relativistic heavy-ion collisions. Regarding conserved charge fluctuations, diffusion plays a vital role as the time evolution of conserved charges can be caused and affected by the diffusion process~\cite{Asakawa:2000wh,Jeon:2000wg,Shuryak:2000pd}. The inelastic interactions do not change the fluctuation because chemical reactions do not change the net charge. But, it is the diffusion process that affects the conserved charge fluctuations. Thus, the study of diffusion processes is essential in the hadronic phase, which can also tell us about any potential trace of the critical point in the conserved charge fluctuations.

In non-relativistic systems, the diffusion process is described by Fick’s law, which relates the diffusion current ($\Delta \vec J_{q}$) corresponding to charge ($q$) originating from the spatial inhomogeneity of the related charge density ($n_{q}(t,\vec x)$): $\Delta \vec J_{q} = -D_{q} \vec \nabla n_{q}(t,\vec x)$. The diffusion coefficient ($D_{q}$) is a dissipative transport coefficient that characterizes the reaction strength of this gradient force. Considering the baryon number fluctuation associated with the baryon number conservation, the diffusion effect is expected to be minimal due to the almost vanishing net baryon density at the mid-rapidity region in ultra-relativistic heavy-ion collisions at RHIC and the LHC. However, as collision energy decreases, the net baryon density increases, leading diffusion processes to play an increasingly important role in the dissipative dynamics of the evolution of hot and dense matter. Due to the relativistic nature of the QCD system under consideration, the non-relativistic Fick’s law gets modified to a relativistic one.  It is noteworthy that, the non-relativistic diffusion coefficient $D_{q}$ and the relativistic diffusion coefficient $\kappa_{q}$ are related by the relation;
$\kappa_{q} = \frac{\partial n_{q}}{\partial \alpha_{q}}D_{q}$, where $\alpha_{q}$ is the thermal potential, given as $\alpha_{q} = \mu_{q}/T$ ~\cite{Fotakis:2019nbq}. Since strongly interacting particles, e.g., hadrons and quarks, can carry more than one conserved charge, the diffusion current of each charge will no longer depend exclusively on the gradient of that particular charge. Since the gradients of every single charge density can generate a diffusion current of any other charge, the diffusion currents of the conserved charges are thus coupled to each other, as proposed by Greif et. al.~\cite{Greif:2017byw}. Therefore, in the presence of multiple conserved charges, one has a generalized Fick’s law with a diffusion matrix, which can be written as,

\[
\begin{pmatrix}
J_{\rm B}^{\nu} \\
\\
J_{\rm Q}^{\nu} \\
\\
J_{\rm S}^{\nu} \\
\end{pmatrix}
=
\begin{pmatrix}
\kappa_{\rm BB} & \kappa_{\rm BQ} & \kappa_{\rm BS} \\
\\
\kappa_{\rm QB} & \kappa_{\rm QQ} & \kappa_{\rm QS} \\
\\
\kappa_{\rm SB} & \kappa_{\rm SQ} & \kappa_{\rm SS} \\
\end{pmatrix}
\cdot
\begin{pmatrix}
\Delta^{\nu} \alpha_{\rm B} \\
\\
\Delta^{\nu} \alpha_{\rm Q} \\
\\
\Delta^{\nu} \alpha_{\rm S} \\
\end{pmatrix}
\]

The diagonal terms of the diffusion matrix give the usual baryon, electric, and strangeness diffusion coefficients (\(\kappa_{\rm BB}, \kappa_{\rm QQ},\) and \(\kappa_{\rm SS}\)). Moreover, the Onsager's theorem dictates that the off-diagonal terms are symmetric. Thus, one can have six unique diffusion coefficients corresponding to the three conserved charges.

Recently, much attention has been given to the estimation of diffusion matrix coefficients. In Ref.~\cite{Greif:2017byw}, the authors have demonstrated that the diffusion currents are coupled. For the first time, they have estimated the diffusion matrix for the hot and dense nuclear matter. The authors chose a hadron resonance gas model for the hadronic sector, and they used a simple quasiparticle model for the partonic sector. In Ref.~\cite{Fotakis:2019nbq}, a hadronic system is studied in a simple (1 + 1)D fluid dynamics approach, where the authors have estimated the diffusion matrix using the kinetic theory. Similarly, the diffusion coefficient has also been calculated in the dynamic quasi-particle model~\cite{Fotakis:2021diq}. In addition, in Ref.~\cite{Das:2021bkz}, the authors have used a hadron resonance gas model by solving the Boltzmann transport equation in the kinetic theory approach and estimated the diffusion matrix. 

The hadron resonance gas model (HRG) describes the hadronic abundances at the chemical freeze-out boundary~\cite{Andronic:2005yp}. In addition, the HRG model explains the lQCD data very well in the low-temperature regime ($T \lesssim 150$ MeV)~\cite{Borsanyi:2012,HotQCD:2012fhj,Bellwied:2015lba,Bellwied:2013cta}. The HRG model assumes all the hadrons as point-like particles with no interactions between them. The thermodynamic properties of strongly interacting matter, such as pressure, energy density, entropy density, and squared speed of sound, have been estimated from the HRG model. A good agreement can be seen with lQCD. However, the HRG model cannot explain the higher-order conserved number susceptibilities estimated by lQCD at higher temperatures~\cite{Bazavov:2013dta}. Thus, modifications have been made to the ideal HRG model. One significant modification is the excluded volume effect in the excluded volume hadron resonance gas (EVHRG) model~\cite{Braun-Munzinger:1999hun}. In this model, the hadrons are assumed to have a finite volume, which acts as an outward pressure in the system. EVHRG explains the lQCD data better than ideal HRG, even explaining the conserved charge fluctuations to some extent~\cite{Bhattacharyya:2013oya}. Recently, another important modification has been made to the ideal HRG model, which included the van der Waals-type interactions~\cite{Vovchenko:2016rkn}. In the van der Waals hadron resonance gas (VDWHRG) model, both attractive and repulsive interactions are considered. It has been shown that the VDWHRG explains the lQCD data better than both ideal HRG and EVHRG~\cite{Vovchenko:2016rkn,Samanta:2017yhh}. The range of agreement with lQCD extends up to $T \simeq 180$ MeV. The VDWHRG model has also been used to estimate various thermodynamic and transport properties and agrees well with other phenomenological models~\cite{Pradhan:2022gbm,Sahoo:2023vkw,Pradhan:2023rvf}.

Moreover, recent studies regarding the possible charm fluctuation have opened up new avenues of understanding the hot and dense nuclear matter. Previously, the charm quantum number was neglected in the diffusion matrix studies with the argument that charm diffusion will be minimal and thus negligible. While it is undoubtedly true that the charm charge diffusion will be less than other charges, the diffusion terms may not necessarily be negligible. In ref.~\cite{Bazavov:2014yba}, the authors have shown significant charm fluctuation in the medium by taking the lattice QCD approach. In one of our previous work~\cite{Goswami:2023hdl}, we have demonstrated the importance of van der Waals interaction in explaining the lQCD data in the charm sector. Furthermore, recent experimental results on open and hidden charmed hadron collectivity suggest that charm thermalization is a strong possibility~\cite{ ALICE:2017quq, CMS:2017vhp, ATLAS:2017xqp, ALICE:2020pvw}. Apart from this, many theoretical studies explore the idea of the thermalized charm sector in heavy-ion collision~\cite{He:2021zej, Wu:2023djn}. Hence, it is important to understand the charm fluctuation and diffusion in the hot and dense medium.
In the QGP phase, the presence of resonance states such as $D$ and $B$ mesons can significantly reduce the relaxation time ($\tau_c$ and $\tau_b$) of heavy quarks dominantly due to collisional energy loss at the low momentum, reducing $\tau_c$ of charm quark up to 2--4 fm at very high temperatures~\cite{vanHees:2004gq, Goswami:2022szb}. Now, in the hadronic phase, the gradient force is relatively weak. As a result, at the low momentum, the Knudsen number of heavy-flavored hadrons such as $D^0$ meson reduced well below one. Therefore, in this paper, we treat charm on the same footing as the other conserved charges, and the same methodology is used to estimate the diffusion of all the conserved charges.         

Thus, in this work, we have considered the charm charge along with the baryon, electric, and strange charge while estimating the diffusion matrix. We calculated the diffusion coefficients using the relaxation time approximation (RTA) method in the van der Waals hadron resonance gas model.
In the presence of charm conserved charge, the modified Fick's law is given by, 
 
\begin{equation}
\begin{pmatrix}
J_{\rm B}^{\nu} \\ 
\\
J_{\rm Q}^{\nu} \\
\\
J_{\rm S}^{\nu} \\
\\
J_{\rm C}^{\nu}
\end{pmatrix}
=
\begin{pmatrix}
\kappa_{\rm BB} & \kappa_{\rm BQ} & \kappa_{\rm BS} & \kappa_{\rm BC} \\
\\
\kappa_{\rm QB} & \kappa_{\rm QQ} & \kappa_{\rm QS} & \kappa_{\rm QC} \\
\\
\kappa_{\rm SB} & \kappa_{\rm SQ} & \kappa_{\rm SS} & \kappa_{\rm SC} \\
\\
\kappa_{\rm CB} & \kappa_{\rm CQ} & \kappa_{\rm CS} & \kappa_{\rm CC} \\
\end{pmatrix}
.
\begin{pmatrix}
\Delta^{\nu} \alpha_{\rm B} \\ 
\\
\Delta^{\nu} \alpha_{\rm Q} \\
\\
\Delta^{\nu} \alpha_{\rm S} \\
\\
\Delta^{\nu} \alpha_{\rm C}
\end{pmatrix} \nonumber
\end{equation}

The article is organized as follows. In section \ref{formulation}, we give a brief formulation of the van der Waals hadron resonance gas along with the estimation of diffusion coefficients within the relaxation time approximation. In Section~\ref{results_discussion}, we discuss the results, and finally, in Section~\ref{summary}, we summarize our findings.

\section{Formulation}
\label{formulation}

\subsection{van der Waals hadron resonance gas model (VDWHRG)}
\label{formulation:vdw}

The ideal Hadron Resonance Gas (HRG) model is a statistical approach that assumes thermal and chemical equilibrium, with non-interacting, point-like hadrons. This model effectively reproduces various thermodynamic properties that align with lattice Quantum Chromodynamics (lQCD) calculations \cite{HotQCD:2014kol}. Furthermore, the ideal HRG model can be used for higher baryochemical potential regions, where the lQCD method faces limitations due to the fermion sign problem~\cite{HotQCD:2014kol, Borsanyi:2013bia}. However, near the critical temperature, some discrepancies arise between the HRG model and lQCD data. This can be attributed to the increasing significance of strong interaction effects as the temperature nears $T_{c}$.

The van der Waals HRG model (VDWHRG) incorporates both attractive and repulsive interactions by introducing the parameters $a$ and $b$, respectively. The model accounts for both attractive and repulsive interactions between pairs of baryons (or antibaryons) but considers only repulsive interactions among mesons~\cite{Vovchenko:2016rkn, Sarkar:2018mbk}. Furthermore, the attractive interactions between mesons are taken care of by considering the higher resonance as stable particles in the HRG model \cite{Pal:2021qav, Dashen:1969ep}. Additionally, short-range interactions between baryons and antibaryons are neglected, as these are dominated by annihilation processes~\cite{Andronic:2012ut}. Due to number fluctuations, the system formed in a relativistic heavy-ion collision closely resembles a grand canonical ensemble (GCE). In the ideal HRG model, the grand canonical partition function for the $i^{th}$ hadronic species is given by \cite{Andronic:2012ut},
\begin{equation}
    \ln Z^{id}_i = \pm \frac{Vg_i}{2\pi^2} \int_{0}^{\infty} p^2 dp\ \ln\{1\pm \exp[-(E_i-\mu_i)/T]\},
\end{equation}
Here, $g_i$, $E_i$, and $\mu_i$ denote the degeneracy, energy, and chemical potential of the $i^{th}$ hadron, respectively. The $\pm$ corresponds to baryon and mesons, owing to Fermi-Dirac and Bose-Einstein distributions. The energy of the $i^{th}$ hadron with mass $m_i$ is given by $E_i = \sqrt{p^2 + m_i^2}$, and the chemical potential $\mu_i$ can be further expressed in terms of the baryon, strangeness, charge, and charm chemical potentials, along with their corresponding conserved quantum numbers, as follows:
\begin{equation}
    \mu_i = B_i\mu_B + S_i\mu_S + Q_i\mu_Q + C_i\mu_C,
\end{equation}
where $B_{i}$, $S_{i}$, $Q_{i}$, and $C_{i}$ are, respectively, the baryon number, strangeness, electric charge, and charm quantum number of the $i^{th}$ hadron. In the ideal HRG formalism, the pressure $P^{id}_{i}$ due to $i^{th}$ hadron in an ideal hadron gas in the GCE can be written as,
\begin{equation}
    P^{id}_i(T,\mu_i) = \pm \frac{Tg_i}{2\pi^2} \int_{0}^{\infty} p^2 dp\ \ln\{1\pm \exp[-(E_i-\mu_i)/T]\},
\end{equation}
The number density $n^{id}_{i}$ and energy density $\epsilon_{i}$ for the $i^{th}$ hadron is  given as,
\begin{equation}
    n^{id}_i(T,\mu_i) = \frac{g_i}{2\pi^2} \int_{0}^{\infty} \frac{p^2 dp}{\exp[(E_i-\mu_i)/T]\pm1},
\end{equation}
\begin{equation}
    \epsilon^{id}_i(T,\mu_i) = \frac{g_i}{2\pi^2} \int_{0}^{\infty} \frac{E_{i}p^2 dp}{\exp[(E_i-\mu_i)/T]\pm1}.
\end{equation}

The equation of state in the VDWHRG formalism is given as,
\begin{equation}
    \label{eqVDW}
    \left( P + a\left( \frac{N}{V} \right)^2 \right)(V - bN) = NT,
\end{equation}
where $P$, $N$, $V$, and $T$ are the pressure, number of particles, volume, and temperature of the system, respectively, and $a$ and $b$ are the van der Waals parameters, where $b=\frac{16}{3}\pi r^{3}$, $r$ being the hardcore radius of the hadron.
In this work, the parameters are set as $a = 0.926$ GeV fm$^{3}$. For the repulsive parameter $b$, the hardcore radius is taken as $r_{M} = 0.2$ $\rm fm$ for mesons and $r_{B(\bar{B})} = 0.62$ $\rm fm$ for baryons (and antibaryons)~\cite{Sarkar:2018mbk}. Here, $b_{M}$ denotes the excluded volume for mesons, and $b_{B(\bar{B})}$ for baryons and antibaryons.

Pressure in Eq.~(\ref{eqVDW}) can be expressed in terms of number density, $n = N/V$ as, 
\begin{equation}
    P(T,n) = \frac{nT}{1-bn} - an^{2}.
\end{equation}
In the GCE, we can express pressure as~\cite{Vovchenko:2015vxa, Vovchenko:2015pya},
\begin{equation}
    P(T,\mu) = P^{id}(T, \mu^{*}) - an^{2}(T,\mu),
    \label{eq_prs_vdw}
\end{equation}
where $n$ is the number density calculated within the VDWHRG model and $\mu^{*}$ is the effective chemical potential, both given by,
\begin{eqnarray}
    n(T,\mu) = \frac{\sum_{i}n_{i}^{id}(T,\mu^{*})}{1+b\sum_{i}n_{i}^{id}(T,\mu^{*})},
    \label{eq_nvdw}\\
    \mu^{*} = \mu - bP(T,\mu) - abn^{2}(T,\mu) + 2an(T,\mu).
\end{eqnarray}
The energy density $\epsilon$ is given as,
\begin{equation}
    \epsilon(T,\mu) = \frac{\sum_{i}\epsilon_{i}^{id}(T,\mu^{*})}{1+b\sum_{i}n_{i}^{id}(T,\mu^{*})} - an^{2}(T,\mu),
     \label{eq_nrg_vdw}
\end{equation}
and total pressure in the VDWHRG model can be written as
\begin{equation}
    P(T,\mu) = P_{M}(T,\mu) + P_{B}(T,\mu) + P_{\bar{B}}(T,\mu).
\end{equation}

Here, $P_{M}(T,\mu)$, $P_{B}(T,\mu)$, and $P_{\bar{B}}(T,\mu)$ represent the pressures of the three subsystems in a Van der Waals hadron gas, i.e. mesons with repulsive interactions, baryons, and antibaryons with Van der Waals interactions, respectively. The pressure of these subsystems can be further expressed as follows:
\begin{eqnarray}
    P_{M}(T,\mu) = \sum_{i\in M}P_{i}^{id}(T,\mu^{*}_{M}),\\
    P_{B}(T,\mu) = \sum_{i\in B}P_{i}^{id}(T,\mu^{*}_{B})-an^{2}_{B}(T,\mu),\\
    P_{\bar{B}}(T,\mu) = \sum_{i\in \bar{B}}P_{i}^{id}(T,\mu^{*}_{\bar{B}})-an^{2}_{\bar{B}}(T,\mu),
\end{eqnarray}
where $M$, $B$, and $\bar{B}$ represent mesons, baryons, and antibaryons, respectively. The effective chemical potentials are denoted by $\mu^*_{M}$ for mesons and $\mu^*_{B(\bar{B})}$ for baryons (antibaryons). Assuming vanishing chemical potentials corresponding to electric charge, strangeness, and charm quantum numbers, i.e., $\mu_{Q} = \mu_{S} = \mu_{C} = 0$, the modified chemical potentials for mesons and baryons can be expressed as,
\begin{equation}
    \mu^{*}_{M} = - b_{M} P_{M}(T,\mu),
\end{equation}
\begin{equation}
    \mu^{*}_{B(\bar{B})} = \mu_{B(\bar{B})}- b_{B(\bar{B})} P_{{B(\bar{B})}}(T,\mu) - ab_{B(\bar{B})}n^{2}_{B(\bar{B})} + 2an_{B(\bar{B})},
\end{equation}
where $n_{M}$ and $n_{B(\bar{B})}$ are the number densities of mesons and baryons (antibaryons) in a VDW hadron gas, given by Eq.~(\ref{eq_nvdw}).

Using the above VDWHRG formalism, we estimate the diffusion matrix coefficients in an interacting hadron gas in the following section.

Moreover, along with the strangeness and charm neutrality condition, we take care of the electric charge ($Q$) to baryon number ($B)$ ratio, $n_Q/n_B\sim$ 0.4, which is assumed to be the isospin asymmetry, $Z/A$ of the colliding nucleus~\cite{Poberezhnyuk:2019pxs, Becattini:2000jw}. Here, $n_Q$ is the net charge density of all the hadrons, and $n_B$ is the net baryon density. Moreover, due to rare charm production, we consider the charm-canonical ensemble following the analogy used for the strange-canonical ensemble in Refs.~\cite{Cleymans:1998yb, Begun:2018qkw}. These conditions affect the total chemical potential and the number density, which in turn affects the energy density and pressure of the system. However, the changes in the thermodynamic quantities are small and only become prominent at a very high chemical potential corresponding to low $\sqrt{s_{\rm NN}}$.

\subsection{Diffusion coefficients}
\label{formulation:diff}
The symmetric energy-momentum tensor $T^{\mu\nu}$ in terms of fluid degrees of freedom can be expressed as,
\begin{equation}
T^{\mu\nu}=-P g^{\mu\nu} + \omega u^{\mu} u^{\nu} + \Delta T^{\mu\nu},
\label{E-M Tensor_1}
\end{equation}
where $P$, $\varepsilon$, and $\omega=\varepsilon+P$ are pressure, energy density, and enthalpy, respectively. The fluid velocity, $u^{\mu}$, is normalized as $u_{\mu} u^{\mu}=1$. Moreover, $\Delta T^{\mu\nu}$ is the dissipative term, which in the Landau-Lifshitz frame takes the form\cite{Das:2021bkz, Albright:2015fpa},
\begin{equation}
\Delta T^{\mu\nu}=\eta\left(D^\mu u^\nu+D^\nu u^\mu+\frac{2}{3} \Delta^{\mu\nu}\theta\right)-\zeta\Delta^{\mu\nu}\theta~.
\label{}
\end{equation}
Here, $\eta$ and $\zeta$ are shear and bulk viscosity, respectively. The projection orthogonal to $u^\mu$ is $\Delta^{\mu\nu} = g^{\mu\nu} - u^{\mu}u^{\nu}$, and shorthand notations are $D^{\mu} \equiv \partial^{\mu} - u^{\mu}D$, $D \equiv u^{\mu}\partial_{\mu}$,  $\theta  \equiv  \boldsymbol{\partial} \cdot \boldsymbol{u}$. We define the conserved current $J_{q}^{\mu}$ related to a conserved charge $q$ as\cite{Albright:2015fpa},
\begin{equation}
J_q^{\mu}=n_qu^\mu+\Delta J_q^\mu.
\end{equation}
Where, the dissipative contribution $\Delta J_{q}^{\mu}$ can be written as,
\begin{equation}
\Delta J_q^\mu=\kappa_{qq'}D^\mu \alpha_{q'}.
\label{Eq_delta_J}
\end{equation}
Here, $\alpha_{q'} = \frac{\mu_{q'}}{T}$ is the thermal potential and $\kappa_{qq'}$ is the diffusion coefficient. 
In kinetic theory, the energy-momentum tensor and the current density are expressed in terms of single-particle distribution function as
\begin{eqnarray}
    T^{\mu \nu} &=& \sum_{i=\rm hadrons} \int \frac{d^3p_{i}}{(2\pi)^3}\frac{p^{\mu}_{i}p^{\nu}_{i}}{E_{i}} f_{i}~,
    \label{E-M Tensor_2}
\\
    J^{\mu}_{q} &=& \sum_{i=\rm hadrons} q_{i} \int \frac{d^3p_{i}}{(2\pi)^3} \frac{p_{i}^{\mu}}{E_{i}} f_{i}~.
    \label{current}
\end{eqnarray}
We expand the total single-particle distribution function $f_i$ according to the Chapman-Enskog expansion~\cite{Chapman_book}, $f_i = f_i^{0} + f_i^{1}+ \mathcal{O}(Kn)$. $f_i^0$ is the equilibrium distribution function. Here, the Knudsen number ($Kn$) is the ratio of the particle's mean free path ($\lambda_{mfp}$) to the characteristic length of the system. The characteristic length is related to the gradient force. Truncation of the series at the first order is justified for light flavors. Though $\lambda_{mfp}$ of heavy flavor is relatively high, for the relatively small gradient in the hadronic phase, the characteristic length would be sufficient to truncate the expansion at the first order of $Kn$ for charmed hadrons at the low momentum.
Therefore total distribution function for $i^{th}$ hadron species can be expressed as
\begin{equation}
    f_{i} = f_{i}^{0}\left(1 + \Phi_{i}\right).
    \label{total_dis_fn}
\end{equation}
The $\Phi_{i}$ represents the contribution from the non-equilibrium dynamics in the system and modifies $T^{\mu \nu}$ and $J^{\mu}_{q}$ by adding the terms $\Delta T^{\mu \nu}$ and $\Delta J^{\mu}_{q}$, respectively. Thus, $\Phi_{i}$ should have the same tensor structure as $\Delta T^{\mu \nu}$ and $\Delta J^{\mu}_{q}$~\cite{Das:2021bkz, Albright:2015fpa},
\begin{equation}
    \Phi_{i} = -A_{i}\theta - \sum_{i}B^{q}_{i}p^{\mu}_{i}D_{\mu}\alpha_{q} + C_{a}p^{\mu}_{a}p^{\mu}_{b}\Sigma_{\mu \nu}~.
    \label{phi}
\end{equation}
Here, $\Sigma_{\mu \nu} \equiv D_{\mu}u_{\nu} + D_{\nu}u_{\mu} -\frac{2}{3}\Delta_{\mu \nu}\theta$. The gradient forces, first and third terms on the right side of Eq.~(\ref{phi}) contribute to bulk~\cite{Chakraborty:2010fr} and shear viscosities~\cite{Dey:2019axu}, respectively. 
Using the second term, the dissipative part of the current, $\Delta J^{\mu}_{q}$ can be obtained from Eq.~(\ref{current}) as,
\begin{equation}
    \Delta J^{\mu}_{q} = \sum_{i,q'} q_{a} \int\frac{d^3p_{i}}{\left ( 2\pi \right)^3}\frac{p_{i}^2}{3E_{i}}f^{0}_{i}B^{q'}_{i}D^{\mu}\alpha_{q'}.
\end{equation}
Comparing the above equation with Eq.~(\ref{Eq_delta_J}), we get
\begin{equation}
    \kappa_{qq'} = \sum_{i} q_{i} \int\frac{d^3p_{i}}{\left ( 2\pi \right)^3}\frac{p_{i}^2}{3E_{i}}f^{0}_{i}B^{q'}_{i}.
    \label{Diffusion_matrix_term}
\end{equation}

To find the function $B^{q}_{i}$, we solve the relativistic Boltzmann transport equation, which in the absence of an external force, is given as,
\begin{eqnarray}
    \frac{p^{\mu}_{i}}{E_{i}} \partial_{\mu}f_{i}  &=& C_{i}
\nn\\  
     \Rightarrow p^{\mu}_{i} \partial_{\mu}f_{i}  &=& E_{i}C_{i}~,
    \label{Boltzmann_eq}
\end{eqnarray}
where $C_{i}$ is the collision kernel, $p^{\mu}$ is the four-momentum of the particle, and $E$ is the energy of the particle. Using $\partial_{\mu} = u_{\mu}D + D_{\mu}$, we solve for the left hand side of Eq.~(\ref{Boltzmann_eq})
\begin{equation}
 p^{\mu}_{i}\partial_{\mu}f_{i}^{0} = \left ( E_{i}D + p_{a}^{\mu}D_{\mu} \right) f_{i}^{0} = E_{i}Df_{i}^{0} + p_{a}^{\mu}D_{\mu}f_{i}^{0}.
    \label{LHS_Boltzmann_eq}
\end{equation}
We can express the equilibrium distribution function $f_{i}^{0}$ as, 
\begin{equation*}
    f_{i}^{0} = \frac{g_{i}}{exp[\beta u\cdot p_{i} - \alpha_{i}] + a_{f}}
\end{equation*}
where $a_{f} = 1$ for the Fermi-Dirac equilibrium distribution function and $a_{f} = -1$ for the Bose-Einstein equilibrium distribution function; also, $\beta = 1/T$ is the inverse of the temperature of the considered medium. Hence, we obtain
\begin{equation}
    Df_{i}^{0} = f_{i}^{0} \left( 1 - af_{i}^{0} \right)\left ( -E_{i}D\beta -\beta p^{\mu}_{i}D u_{\mu} + D\alpha_{i} \right)
    \label{Eq_D_f}
\end{equation}
and,
\begin{equation}
    \begin{split}
        p_{i}^{\mu}D_{\mu}f_{i}^{0} = f_{i}^{0}\left( 1 - af_{i}^{0} \right) ( -E_{i}p_{i}^{\mu}D_{\mu}\beta \\ - \beta p_{i}^{\mu}p_{i}^{\nu}D_{\mu}u_{\nu} - p_{i}^{\mu}D_{\mu}\alpha^{i})
    \end{split}
    \label{Eq_D_mu_f}
\end{equation}
Following \cite{Das:2021bkz}, we re-write Eq.~(\ref{Eq_D_f}) in terms of charge number density $n_{q}$ and enthalpy, $\omega = \epsilon + P$ and simplifying Eq.~(\ref{Eq_D_mu_f}), we only consider the diffusion part, i.e. the term containing $D\alpha_{i}$. Thus Eq.~(\ref{LHS_Boltzmann_eq}) simplifies to,
\begin{equation}
    p^{\mu}_{i}\partial_{\mu}f_{i}^{0} = - f_{i}^{0}p_{i}^{\mu}\sum_{q}\left( \frac{E_{i}n_{q}}{\omega} - q_{i} \right)D_{\mu}\alpha_{q}
    \label{Boltzmann_Eqn_LHS_modified}
\end{equation}
Now, for the right-hand side of the Eq.~(\ref{Boltzmann_eq}), we expand the collision term using the Chapman-Enskog approximation,
\begin{equation}
    \begin{split}
        C_{i} = \frac{1}{2}E_{i}f_{i}^{0}\sum_{j,k,l}\int \frac{d^3p_{j}}{(2\pi)^{3}} \frac{d^3p_{k}}{(2\pi)^{3}} \frac{d^3p_{l}}{(2\pi)^{3}} f_{j}^{0} W(a,b|c,d) \\ \times \left ( \Phi_{k} + \Phi_{l} - \Phi_{i} - \Phi_{j} \right )
    \end{split}
\end{equation}
where, $W(a,b|c,d)$ is the transition rate, and considering only the diffusive part, we can write $\Phi_{i} \simeq - \sum_{q}B_{i}^{q}p_{i}^{\mu}D_{\mu}\alpha_{q}$. We compare the coefficient of $D_{\mu}\alpha_{q}$ from Eq.~(\ref{Boltzmann_Eqn_LHS_modified}) to get,
\begin{equation}
    \begin{split}
        \frac{p_{i}^{\mu}}{E_{i}}\left( \frac{E_{i}n_{q}}{\omega} - q_{i} \right) = \frac{1}{2}\sum_{j,k,l}\int \frac{d^3p_{j}}{(2\pi)^{3}} \frac{d^3p_{k}}{(2\pi)^{3}} \frac{d^3p_{l}}{(2\pi)^{3}} f_{j}^{0} \\ \times W(i,j|k,l) \left ( B^{q}_{k}p_{k}^{\mu} + B^{q}_{l}p_{l}^{\mu} - B^{q}_{i}p_{i}^{\mu} - B^{q}_{j}p_{j}^{\mu} \right )
        \label{Eq32}
    \end{split}
\end{equation}
The above equation does not have a unique solution, and this can be observed by substituting $B^{q}_{i}$ as $B^{q}_{i} = B^{q}_{1i} - b^{q}$, where $b^{q}$ is a constant independent of momentum and hadronic species. Thus, further constraints are added by taking the local rest frame as~\cite{Albright:2015fpa}
\begin{equation}
    \Delta T^{0\nu} = 0, \Delta J^{0}_{q} = 0
\end{equation}
We estimate $\Delta T^{0\nu} = 0$ by using Eq.~(\ref{E-M Tensor_2}) and replacing the distribution function $f$ from Eq.~(\ref{total_dis_fn}), where we consider only the diffusion term i.e. $D_{\mu}\alpha_{q}$, we get,
\begin{equation}
    \sum_{q}\sum_{i}\int \frac{d^3p_{i}}{(2\pi)^3}p_{i}^{\mu}B_{i}^{q}p_{i}^{\nu}D_{\nu}\alpha_{q}f_{i}^{0} = 0
\end{equation}
We substitute $B_{i}^{q}$ with $B^{q}_{i} = B^{q}_{1i} - b^{q}$, that gives us
\begin{equation}
    \sum_{q}\sum_{i}\int \frac{d^3p_{i}}{(2\pi)^3}p_{i}^{\mu}\left( B^{q}_{1i} - b^{q} \right )p_{i}^{\nu}D_{\nu}\alpha_{q}f_{i}^{0} = 0
\end{equation}
\begin{equation}
    b_{q}\sum_{i}\int \frac{d^3p_{i}}{(2\pi)^3}p_{i}^{2}f_{i}^{0} = \sum_{i}\int \frac{d^3p_{i}}  {(2\pi)^3}p_{i}^{2}B^{q}_{1i}f_{i}^{0}
\end{equation}
We can express $b^{q}$ as~\cite{Das:2021bkz},
\begin{equation}
    b^{q} = \frac{1}{3T\omega}\sum_{i}\int \frac{d^3p_{i}}{(2\pi)^3}p_{i}^{2}B^{q}_{1i}f_{i}^{0}
    \label{b_q_term}
\end{equation}
Thus, we can use the substitution $B^{q}_{i} = B^{q}_{1i} - b^{q}$ in Eq.~(\ref{Diffusion_matrix_term}) to get,

\begin{equation}
    \begin{split}
        \kappa_{qq'} = \sum_{i} q_{i} \int\frac{d^3p_{i}}{\left ( 2\pi \right)^3}\frac{p_{i}^2}{3E_{i}}B^{q'}_{1i}f^{0}_{i} \\ - b_{q'} \sum_{i} q_{i} \int\frac{d^3p_{i}}{\left ( 2\pi \right)^3}\frac{p_{i}^2}{3E_{i}}f^{0}_{i}
    \end{split}
\end{equation}

For the term multiplied by $b_{q'}$ can be defined as,
\begin{equation}
    \sum_{i,q'} q_{i} \int\frac{d^3p_{i}}{\left ( 2\pi \right)^3}\frac{p_{i}^2}{3E_{i}}f^{0}_{i}\left( p \right) = n_{q} T
    \label{n_q_term}
\end{equation}
where, $n_{q}$ is the net charge density and $T$ is the temperature of the system.
Combining Eq.~(\ref{b_q_term}) and \ref{n_q_term}, we can express $\kappa_{qq'}$ as
\begin{equation}
    \kappa_{qq'} = \sum_{i} \int\frac{d^3p_{i}}{\left ( 2\pi \right)^3}\frac{p_{i}^2}{3E_{i}}\left( q_{i} - \frac{n_{q}E_{i}}{\omega} \right)B^{q'}_{1i}f^{0}_{i}
\end{equation}
For the estimation of $\kappa_{qq'}$, we use relaxation time approximation to find the form of $B^{q'}_{1i}$. In this approximation, we assume that all the particles are in equilibrium except the $i^{th}$ hadronic species. Thus, Eq.~(\ref{Eq32}) reduces to, 
\begin{equation}
    \begin{split}
        \frac{p_{i}^{\mu}}{E_{i}}\left( \frac{E_{i}n_{q}}{\omega} - q_{i} \right) = \frac{1}{2}\sum_{j,k,l}\int \frac{d^3p_{j}}{(2\pi)^{3}} \frac{d^3p_{k}}{(2\pi)^{3}} \frac{d^3p_{l}}{(2\pi)^{3}} \\ \times W(i,j|k,l) \left (- B^{q}_{1i}p_{i}^{\mu} \right)
    \end{split}
\end{equation}
Here, the coefficient of $B^{q}_{1i}$ is the inverse of the relaxation time of the $i^{th}$ hadronic species given as,
\begin{equation}
    \tau^{-1}_{i} = \frac{1}{2}\sum_{j,k,l}\int \frac{d^3p_{j}}{(2\pi)^{3}} \frac{d^3p_{k}}{(2\pi)^{3}} \frac{d^3p_{l}}{(2\pi)^{3}} W(i,j|k,l)
    \label{tau_inv}
\end{equation}
Hence, we can express $B^{q}_{1i}$ in terms of relaxation time as,
\begin{equation}
    B^{q}_{1i} = \frac{\tau_{i}}{E_{i}}\left(q_{i} - \frac{E_{i}n_{q}}{\omega}  \right)
\end{equation}
The final expression for $\kappa_{qq'}$ becomes
\begin{equation}
    \kappa_{qq'} = \sum_{i} \int\frac{d^3p_{i}}{\left ( 2\pi \right)^3}\frac{p_{i}^2}{3E_{i}}\left( q_{i} - \frac{n_{q}E_{i}}{\omega} \right)\frac{\tau_{i}}{E_{i}}\left(q'_{i} - \frac{n_{q'}E_{i}}{\omega}  \right)f^{0}_{i}
    \label{eq:final}
\end{equation}
We use this expression to estimate the diffusion matrix coefficient for all four conserved charges. In this equation, we take $n_{q}$ as the net charge density, i.e., $n_{q} = \sum_{i}q_{i}n_{i}$, where $q_{i}$ is the conserved charge and  $n_{i}$ is the number density of the $i^{th}$ hadron species respectively, estimated using Eq.~(\ref{eq_nvdw}). The enthalpy of the system, $\omega = \epsilon + P$, where Pressure and energy are estimated using Eqs.~(\ref{eq_prs_vdw}) and Eq~(\ref{eq_nrg_vdw}), respectively. For the estimation of relaxation time, we see that Eq.~(\ref{tau_inv}) in the center-of-mass frame can be written as \cite{Das:2021bkz, Kadam:2015xsa}
\begin{equation}
\begin{split}
        \tau^{-1}_{i} &= \sum_{j}\int \frac{d^3p_{j}}{(2\pi)^{3}} \sigma_{ij}\frac{\sqrt{s-4m^2}}{2E_i 2E_j}f^0_j \\ 
        &\equiv \sum_{j}\int \frac{d^3p_{j}}{(2\pi)^{3}} \sigma_{ij}v_{ij}f^0_j,
        \label{}
        \end{split}
\end{equation}
where $\sigma_{ij}$ represents the total scattering cross-section for the process, $i(p_i) + j(p_j)\rightarrow i(p_k) + j(p_l)$. $v_{ij}$ is the relative velocity between $i^{th}$ and $j^{th}$ hadronic species, and $\sqrt{s}$ is the centre-of-mass energy. The above relation gives the energy-dependent relaxation time. However, for simplicity, one can use the average relaxation time as a good approximation \cite{Das:2021bkz, Kadam:2015xsa}. Hence, averaging over $f_i^0$,
\begin{equation}
\begin{split}
        \Tilde{\tau}^{-1}_{i} = \frac{\int \frac{d^3p_{i}}{(2\pi)^{3}}\tau^{-1}_{i}f_i^0}{\frac{d^3p_{i}}{(2\pi)^{3}}f_i^0} = \sum_{j}\frac{\int \frac{d^3p_{i}}{(2\pi)^{3}}\frac{d^3p_{j}}{(2\pi)^{3}}\sigma_{ij}v_{ij}f_i^0f_j^0}{\frac{d^3p_{i}}{(2\pi)^{3}}f_i^0}.\\
\end{split}
\end{equation}
Therefore, the average relaxation time used in this work is given by,
\begin{equation}
    \Tilde{\tau}^{-1}_{i} = \sum_{j} n_{j} \langle \sigma_{ij}v_{ij} \rangle.
    \label{eqtau}
\end{equation}
The thermal average, $\langle \sigma_{ij}v_{ij} \rangle$ is then estimated as,
\begin{equation}
    \begin{split}
            \langle \sigma_{ij} v_{ij} \rangle = \frac{\sigma_{ij}}{8Tm_{i}^{2}m_{j}^{2}K_{2}(\frac{m_{i}}{T})K_{2}(\frac{m_{j}}{T})} \int_{(m_{i} + m_{j})^{2}}^{\infty} ds \\ \times \frac{s-(m_{i}-m{j})^{2}}{\sqrt{s}} (s-(m_{i}+m_{j})^2)K_{1}\Big(\frac{\sqrt{s}}{T}\Big). 
    \end{split}
\end{equation}
Here, $m_{i}$ and $m_{j}$ is the mass of $i^{th}$ and $j^{th}$ hadronic species respectively. $s = (p_{i} + p_{j})^2$ is the Mandelstam variable, and $K_{n}$ is the modified Bessel function of $n$th order.
Moreover, we compute the diagonal component of the diffusion matrix as a function of $\sqrt{s_{\text{NN}}}$. In doing so, we need a relation among the parameters, i.e., T, $\mu_{B}$ and $\sqrt{s_{\text{NN}}}$. We make use of the parametrization given in Ref.\cite{Cleymans:2005xv} as,
\begin{equation*}
    T(\mu_{B}) = q_{1} - q_{2}\mu_{B}^{2} - q_{3}\mu_{B}^{4},
\end{equation*}
\begin{equation*}
    \mu_{B}(\sqrt{s_{\text{NN}}}) = \frac{q_{4}}{1+q_{5}\sqrt{s_{\text{NN}}}}.
\end{equation*}
Here, $q_{1} = 0.166~\text{GeV}$, $q_{2} = 0.139~\text{GeV}^{-1}$, $q_{3} = 0.053~\text{GeV}^{-3}$, $q_{4} = 1.308~\text{GeV}$, and $q_{5} = 0.273~\text{GeV}^{-1}$. These parameters are obtained using freeze-out criteria based on the ideal HRG model. While similar estimations have been made using the HRG model with VDW interactions and the excluded volume HRG model, the variations in the parameters are negligible~\cite{Behera:2022nfn, Tiwari:2011km}.

\section{Results and Discussion}
\label{results_discussion}

\begin{figure*}
    \centering
    \includegraphics[width = 0.45\linewidth]{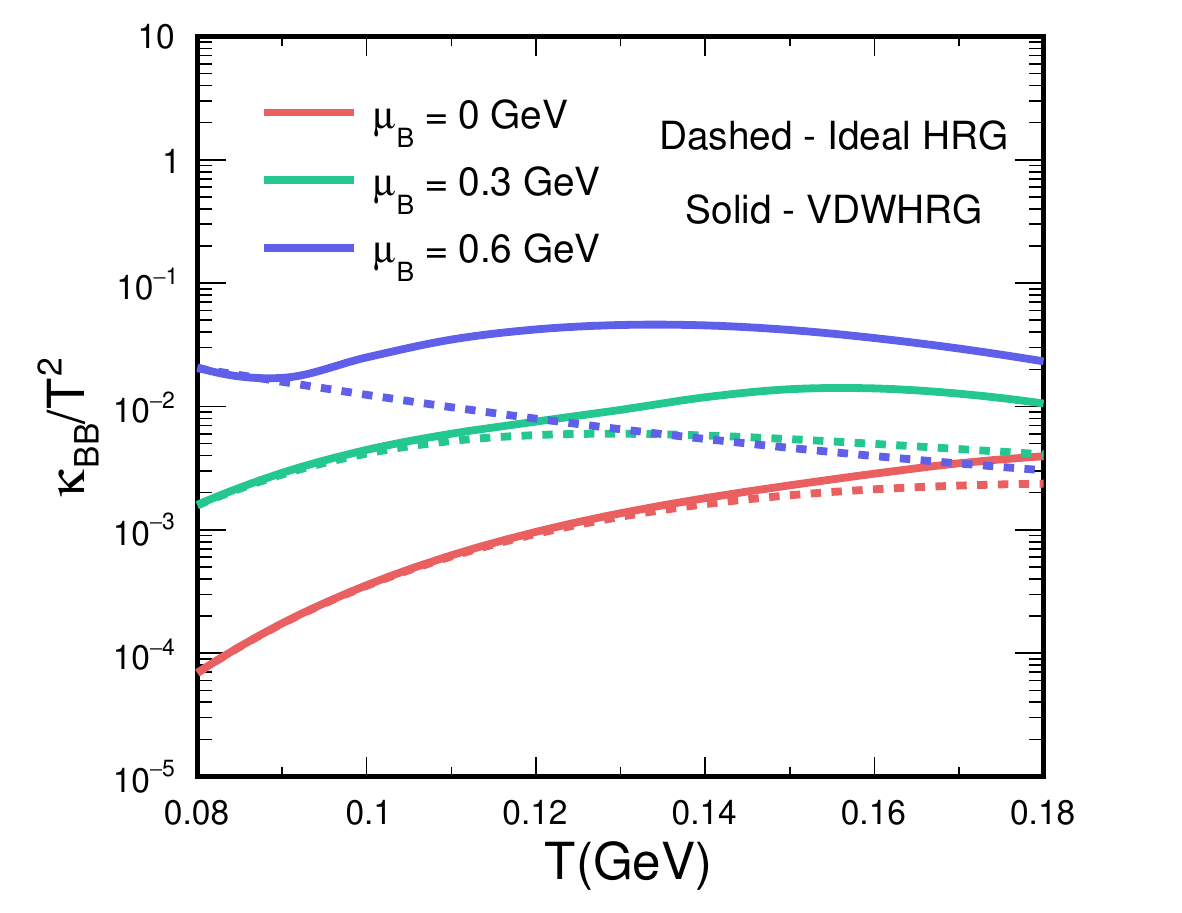}
    \includegraphics[width = 0.45\linewidth]{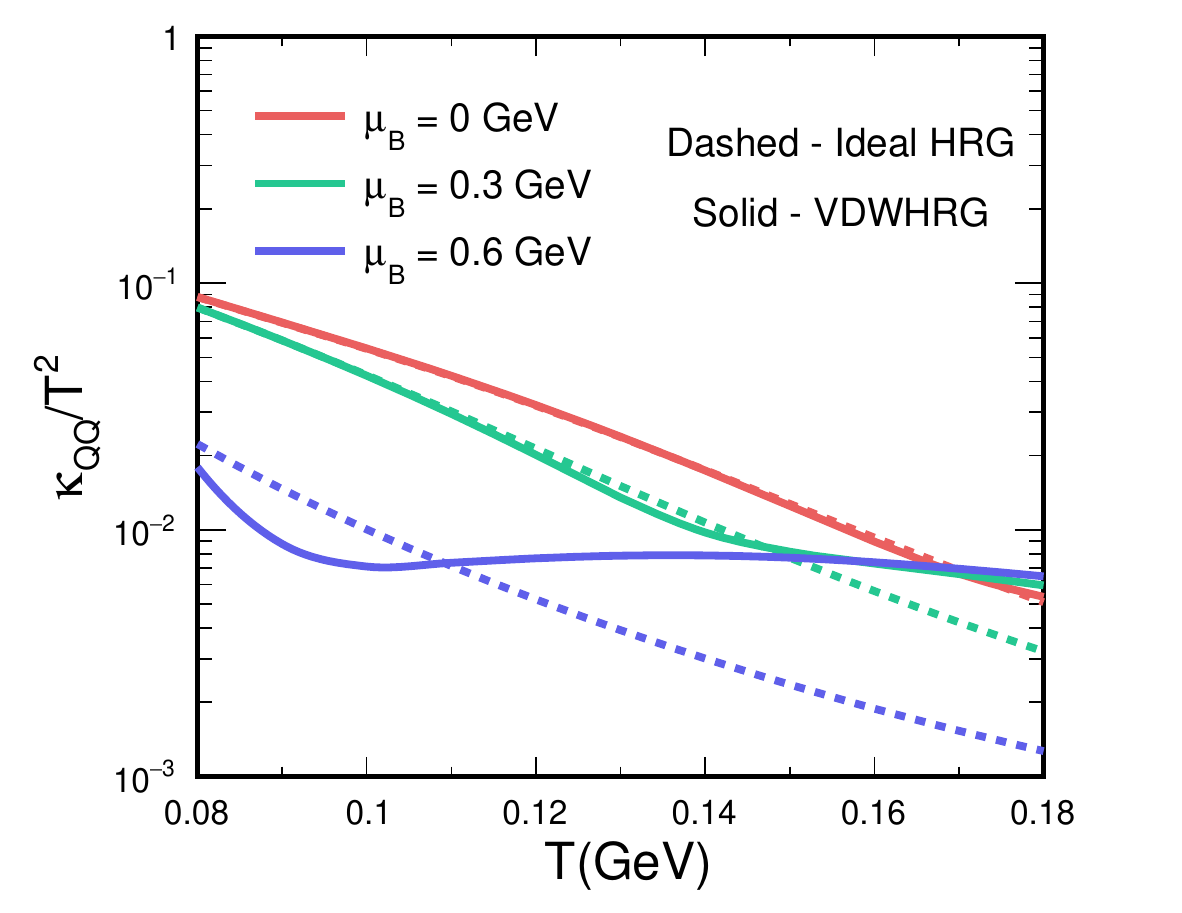}
    \includegraphics[width = 0.45\linewidth]{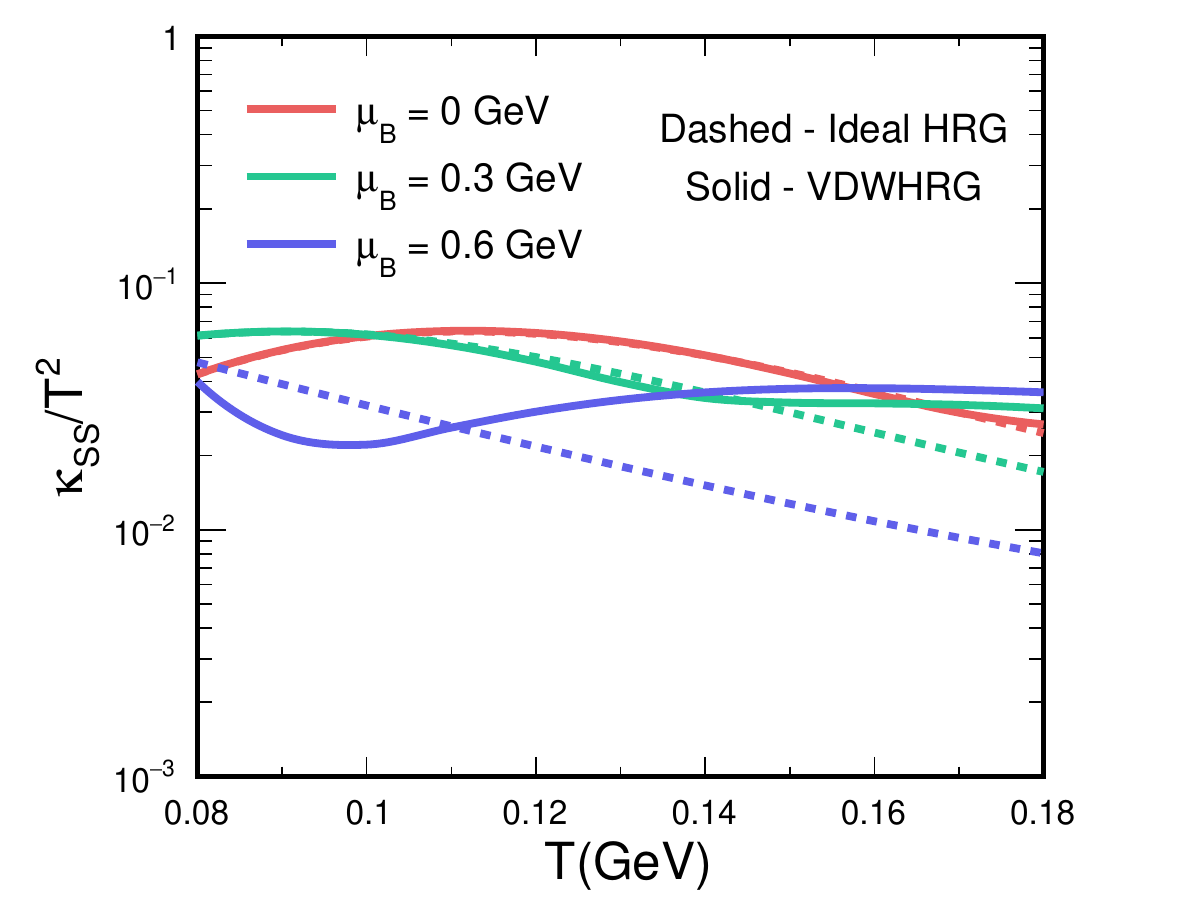}
    \includegraphics[width = 0.45\linewidth]{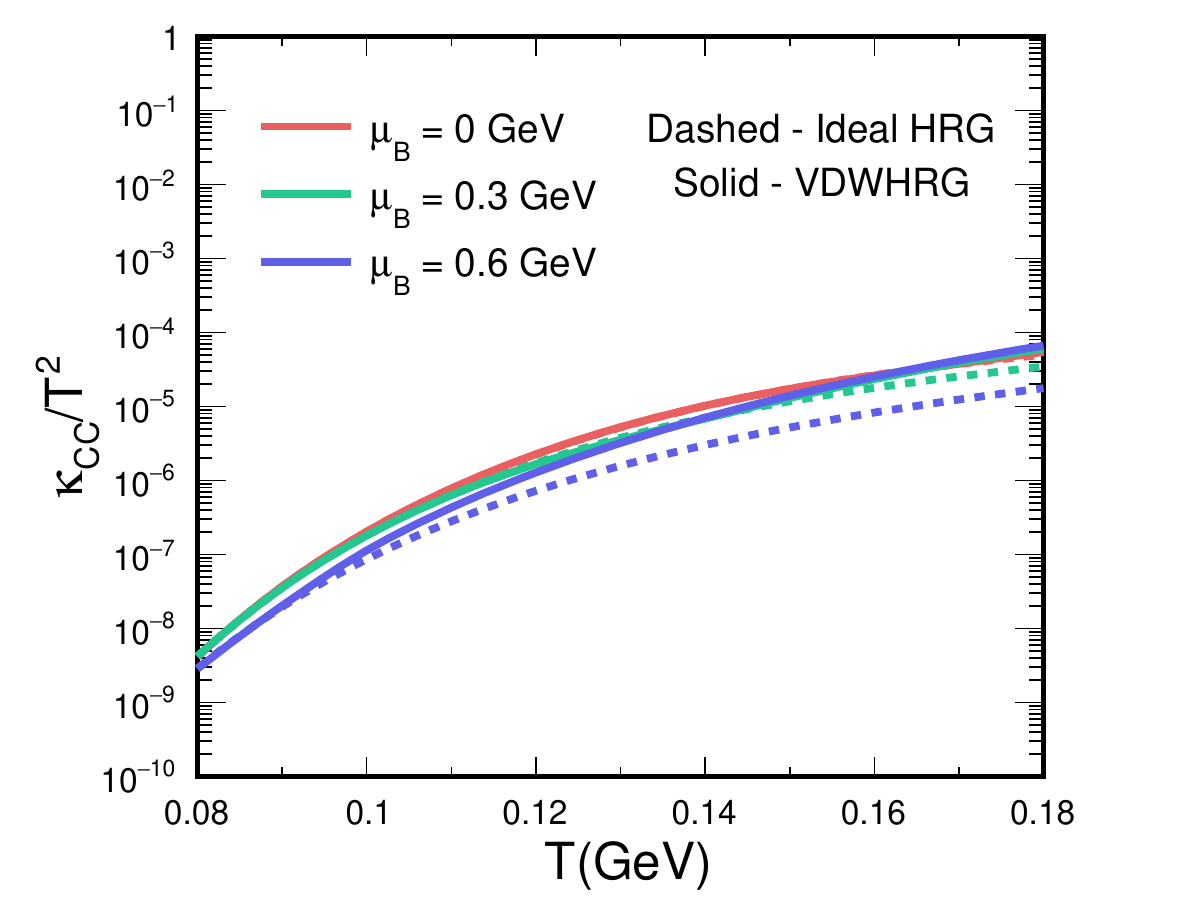}
    
    \caption{The diagonal terms of the diffusion matrix related to the baryon, electric charge, strangeness, and charm charge as a function of temperature for three different baryochemical potentials.}
    \label{fig1}
\end{figure*}

We estimate the diffusion coefficient related to the baryon ($\kappa_{\rm BB}$), charge ($\kappa_{\rm QQ}$), strangeness ($\kappa_{\rm SS}$) and charm number ($\kappa_{\rm CC}$) in a hadronic medium with VDW-like interactions among the hadrons. We primarily use Eq.~(\ref{eq:final}) to estimate the diffusion coefficient matrix components. Figure~\ref{fig1} illustrates the variations in the diagonal components of the diffusion matrix with temperature for three distinct baryochemical potential values. 

To understand the temperature dependency of the diffusion coefficients, we see that Eq.~(\ref{eq:final}) can be divided into three different parts. Firstly, the contribution comes from the relaxation time $(\tau)$, which decreases with temperature. With increasing $T$ the collision rate increases, which reduces $\tau_a$, and for the lighter particles the relative velocity is higher, which further reduces $\tau_a$. Secondly, the contribution comes from the integration of the distribution function, which is similar to the total number density of a hadronic species, which increases with increasing temperature. Lastly, the $\left( \frac{n_{q}E}{\omega} \right)$ term, which vanishes at $\mu_{B}$ = 0 GeV, contributes as the net charge density $n_{q}$ increases as a function of temperature. However, the change in net charge density as a function of temperature is different for electrical charge $(n_{Q})$ as compared to net charge density for the other conserved charges $(n_{B})$, $(n_{S})$, and $(n_{C})$. Moreover, at vanishing chemical potential, the net charge density vanishes. Therefore, at $\mu \to 0$, $\frac{\kappa_{qq}}{T^2} = \frac{\sigma_q}{T}$; so, the dimensionless diffusion coefficient depicts the same behavior as of the conductivity of corresponding charge. In Fig.~(\ref{fig1}), $\kappa_{BB}/T^{2}$ and $\kappa_{CC}/T^{2}$, for $\mu_{B}$ = 0 (solid red line), show an increasing trend as a function of temperature. This indicates that the total number density overcomes the corresponding relaxation time contribution, thus resulting in a net increasing trend. However, this is not the case for $\kappa_{SS}/T^{2}$ and $\kappa_{QQ}/T^{2}$, where $\tau$ dominates, and this explains the decreasing trend for charge and strange diffusion at $\mu_B = 0$. 

For non-vanishing $\mu_{B}$ ($\mu_{Q}$ and $\mu_{S}$) values at low temperature $(T \sim \rm{0.08-0.1 ~GeV})$, the contribution from net charge density, $\left( \frac{n_{q}E}{\omega} \right)$, arises. As $\mu_B$ increases, $\mu_S$ increases with it, however, $\mu_Q$ decreases.
This affects $\kappa_{QQ}/T^{2}$ differently as compared to $\kappa_{SS}/T^{2}$ and $\kappa_{CC}/T^{2}$. The $\kappa_{QQ}/T^{2}$ values decrease at low temperatures as the chemical potential increases from 0.3 to 0.6 GeV. However, the trend is reversed for $\kappa_{BB}/T^{2}$ because of the dominating contribution from total baryon density, which increases massively with increasing $\mu_{B}$.
Moreover, for non-vanishing chemical potential, we observe a general decreasing trend for $\kappa_{SS}/T^{2}$ and $\kappa_{QQ}/T^{2}$ for IHRG results, which can be attributed to the dominance of relaxation time over number density. For $\kappa_{BB}/T^{2}$, we observe that for $\mu_{B}$ = 0.3 GeV, there is an interplay between relaxation time and number density. However, for $\mu_{B}$ = 0.6 GeV, we observe a clear decreasing trend.
On the contrary, for $\kappa_{CC}/T^{2}$, we observe an increasing trend as a function of temperature for all the $\mu_{B}$ values. Such behaviors have already been reported by authors in Ref.~\cite{Das:2021bkz} for all the charges except for the charm case. 

On introducing the VDW-like interactions, an enhancement in results compared to the IHRG case at high $T$ is noticeable. The deviation becomes significant with increasing $\mu_B$. In VDWHRG, number density decreases compared to IHRG, with the relation, $n_{\rm VDW} = \frac{n_{\rm id}}{1+b~n_{\rm id}}$. A decrease in the number density dilutes the gas such that particles can diffuse a longer path without interacting. This leads to higher relaxation time, which in turn increases the $\kappa_{qq'}$ values. The deviation of number density in VDWHRG formalism as compared to IHRG is larger at higher $T$ and $\mu_{B}$. Thus, the deviation in the diagonal components is more prominent in high $T$ and $\mu_B$.

Additionally, the dip in the values of $\kappa_{BB}$, $\kappa_{QQ}$, and $\kappa_{SS}$ is observed near $\mu_{\rm B}$ = 0.6 GeV and $T \sim$ 0.09 GeV due to the occurrence of the liquid-gas phase transition, which culminates in a critical point at low temperatures and high $\mu_{\rm B}$~\cite{Samanta:2017yhh, Sarkar:2018mbk, Vovchenko:2015pya}. Finally, in the bottom-right panel, the charm diffusion is significantly smaller than the other charge diffusion due to their higher masses of charmed hadrons in the hadronic medium. As a result, even though it exists, the liquid-gas phase transition is not apparent in $\kappa_{CC}$ values. Here, it is worthy of pointing out that, as the first work of estimating the diffusion coefficient matrix for the charmed hadrons, we observe a very small magnitude of $\kappa_{CC}/T^{2}$, which indicates that we have a very small charm current. Most importantly, it is not negligible, and at $T \sim$ 0.18 GeV, $\kappa_{CC}/T^{2}$ rises to nearly four orders of magnitude, indicating that temperature is an important factor in the charm diffusion.

\begin{figure}
    \centering
    \includegraphics[width=0.98\linewidth]{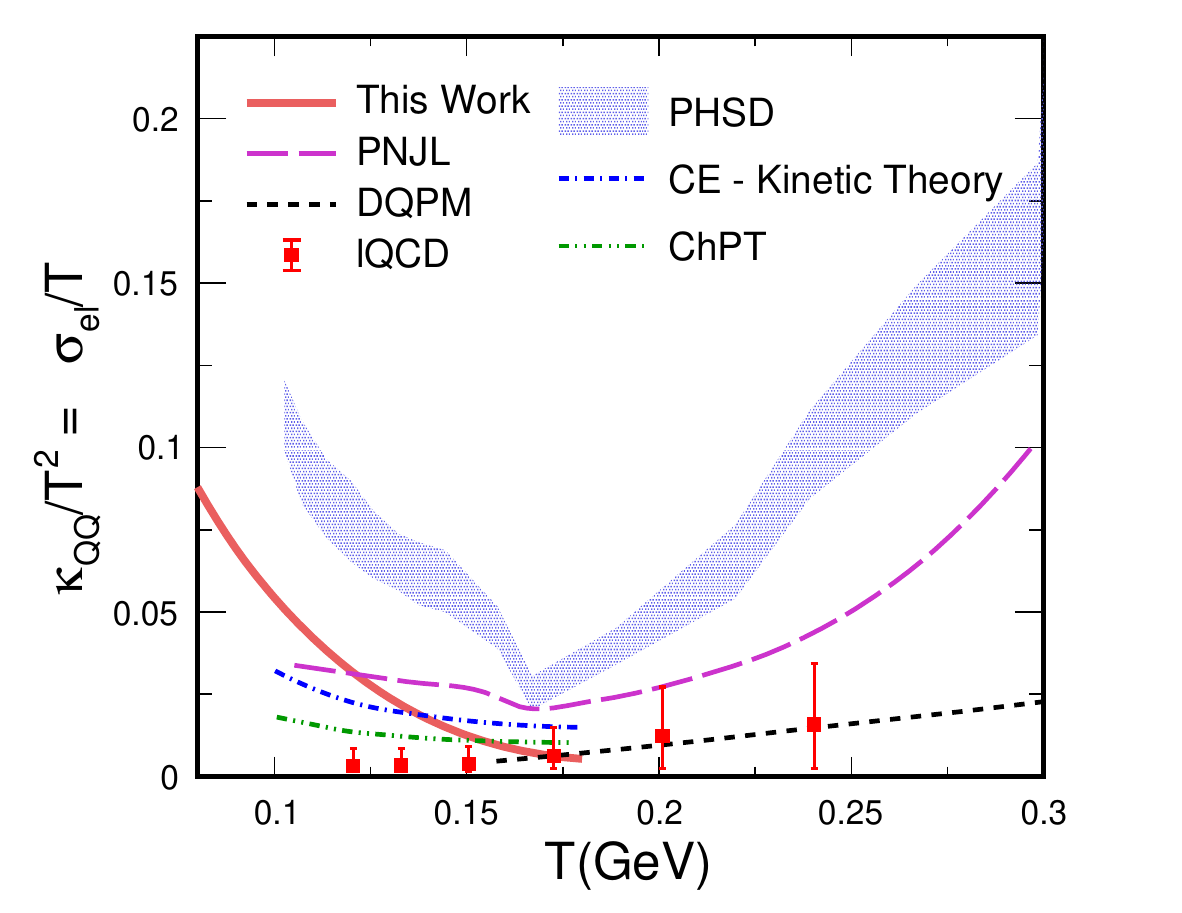}
    \caption{Scaled electrical conductivity as a function of temperature. The blue dashed-dotted line is estimated from kinetic theory~\cite{Greif:2016skc}, and the green dashed-dotted-dotted line shows results from chiral perturbation theory~\cite{Fernandez-Fraile:2005bew}. The magenta dashed line is estimated from the PNJL model~\cite{Soloveva:2020hpr}, and the black dashed line is from DQPM formalism~\cite{Soloveva:2019xph}. The blue band is the result of PHSD calculations~\cite{Cassing:2013iz}. The red squares represent the lQCD data~\cite{Aarts:2014nba}.}
    \label{fig:EC}
\end{figure}

\begin{figure*}

    \centering
    \includegraphics[width = 0.3\linewidth]{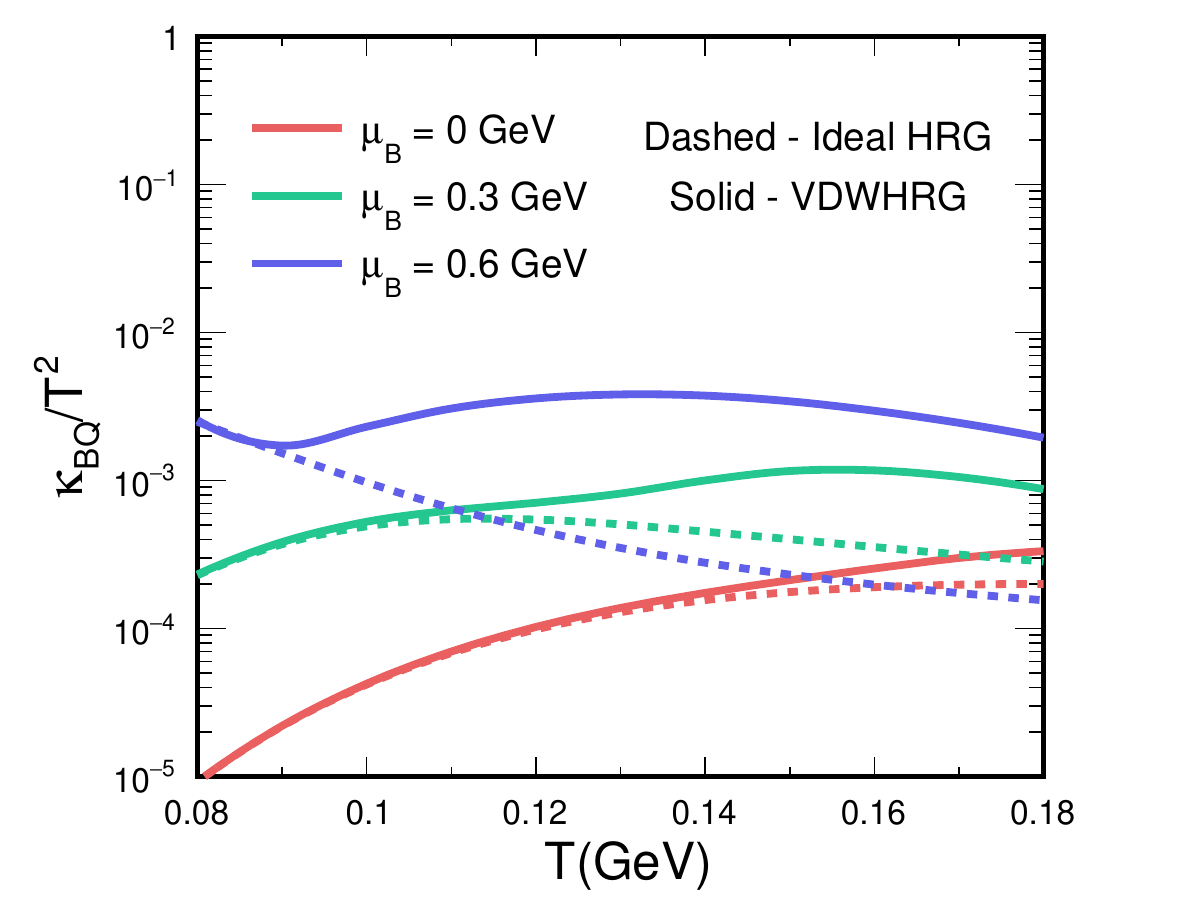}
    \includegraphics[width = 0.3\linewidth]{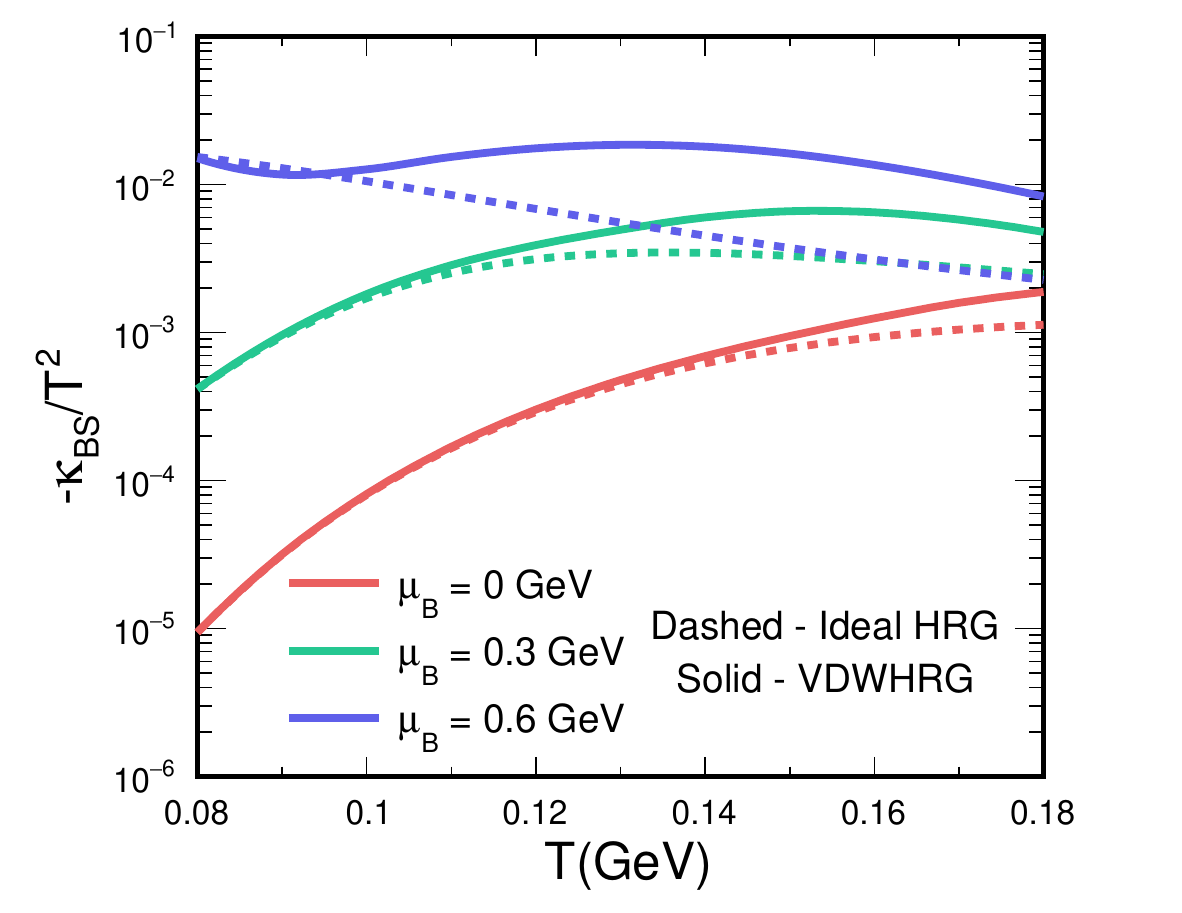}
    \includegraphics[width = 0.3\linewidth]{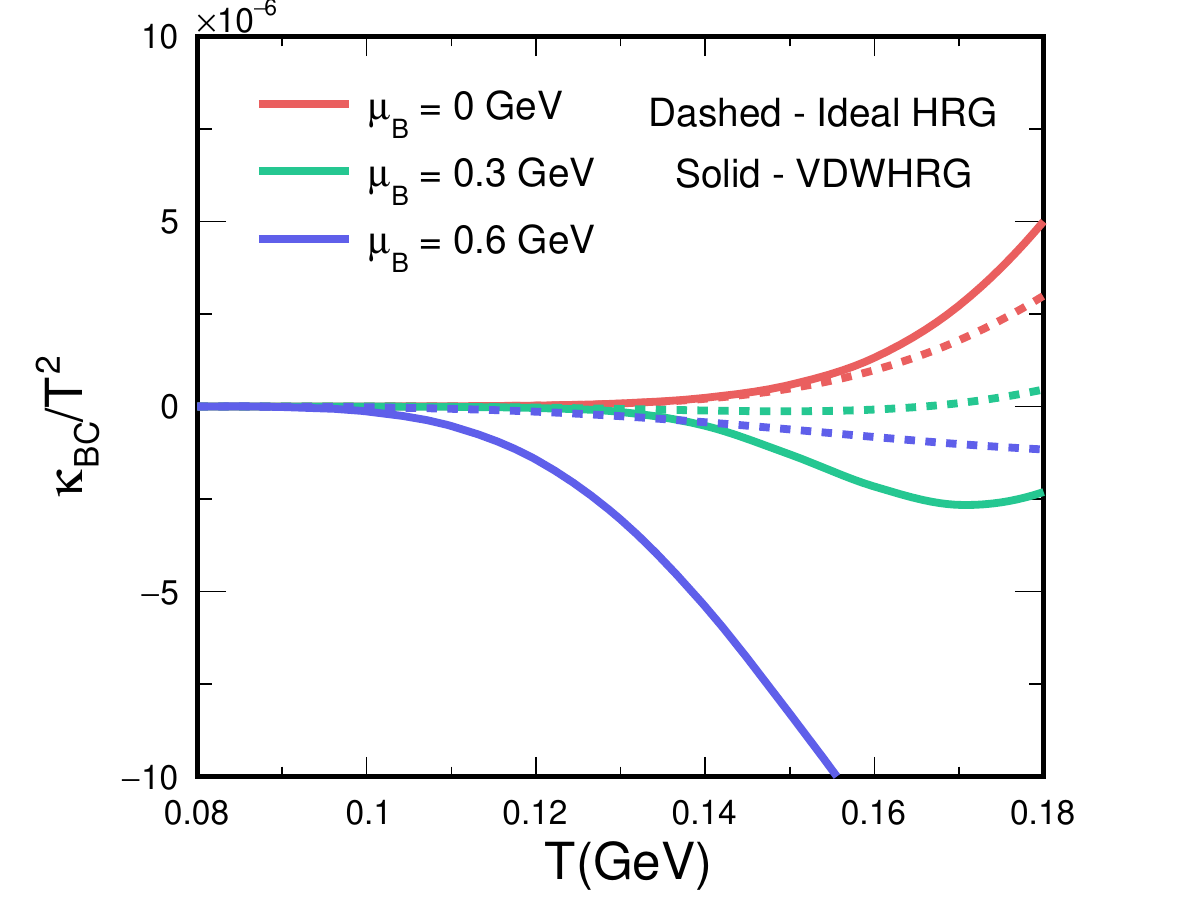}
    \includegraphics[width = 0.3\linewidth]{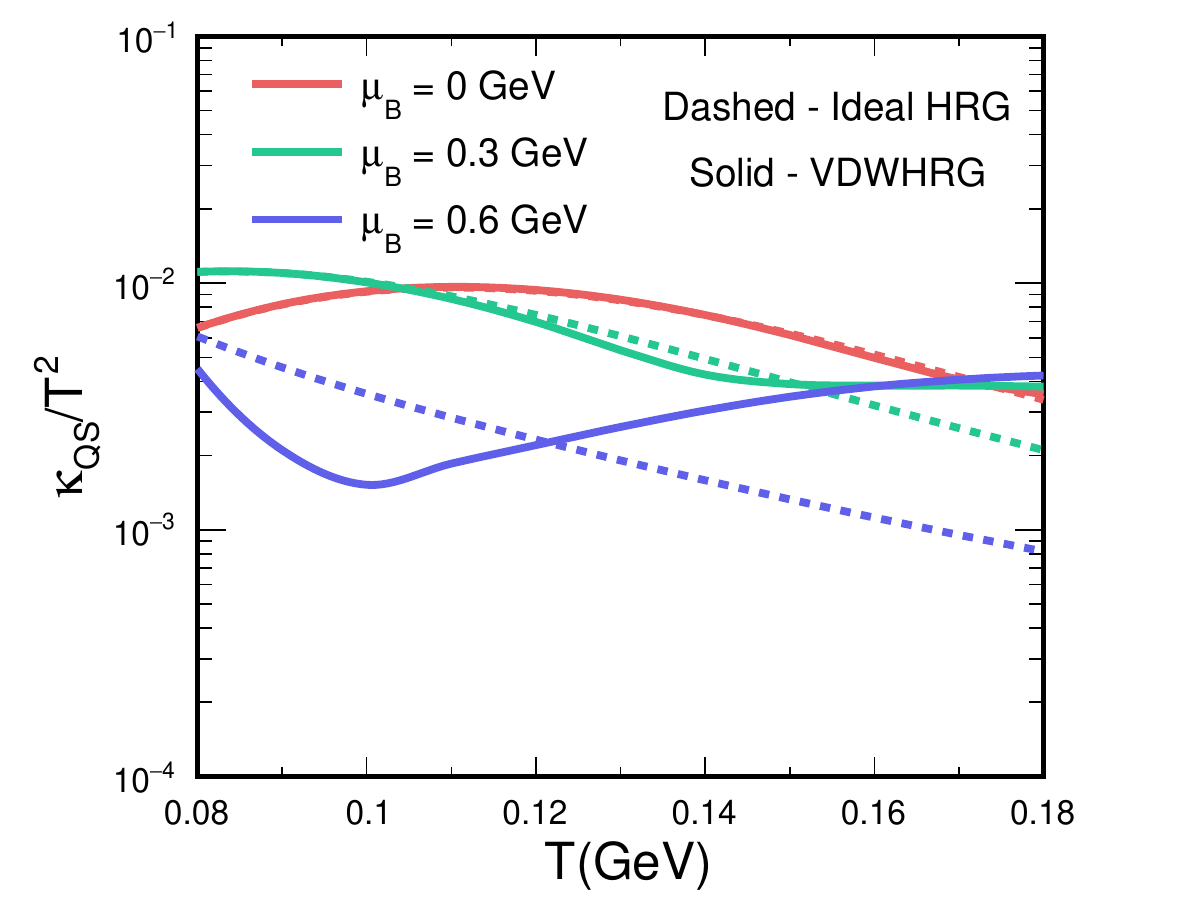}
    \includegraphics[width = 0.3\linewidth]{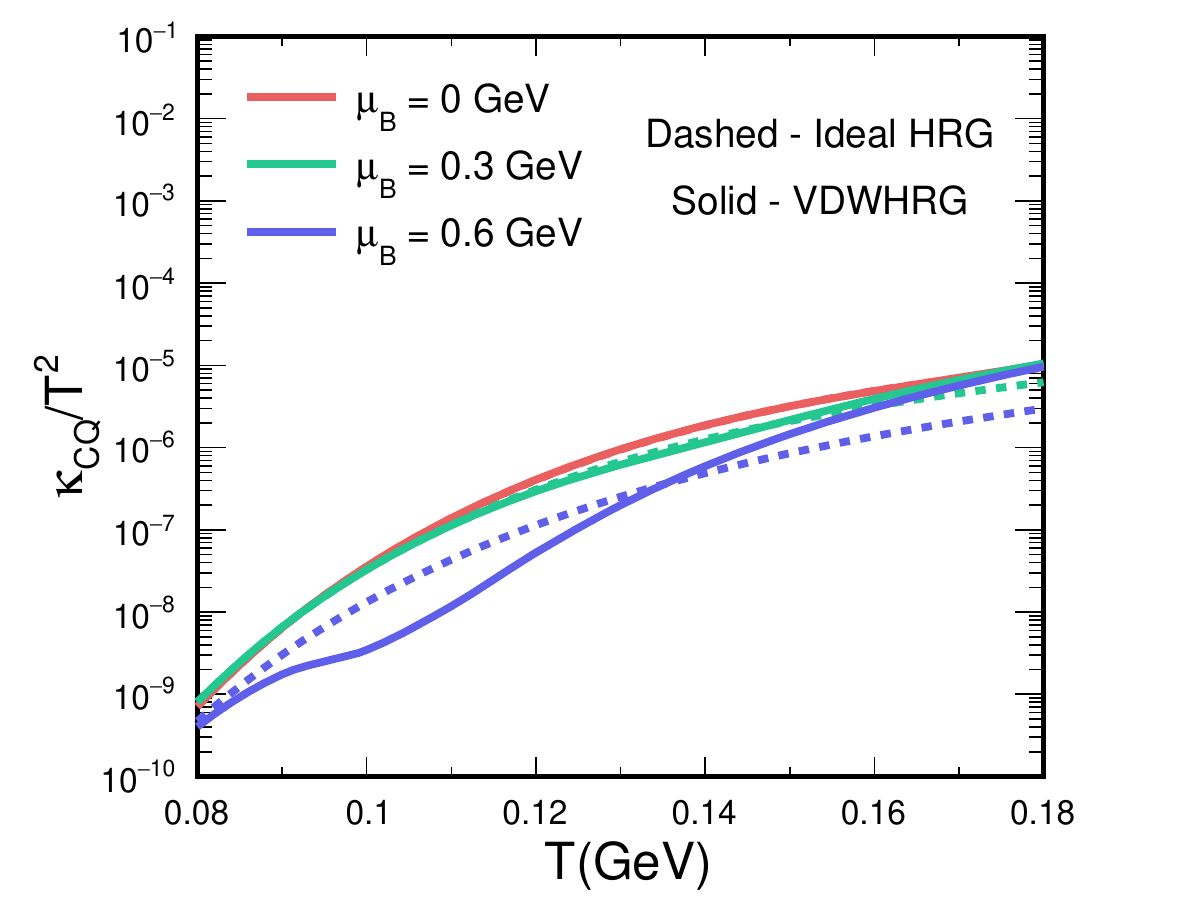}
    \includegraphics[width = 0.3\linewidth]{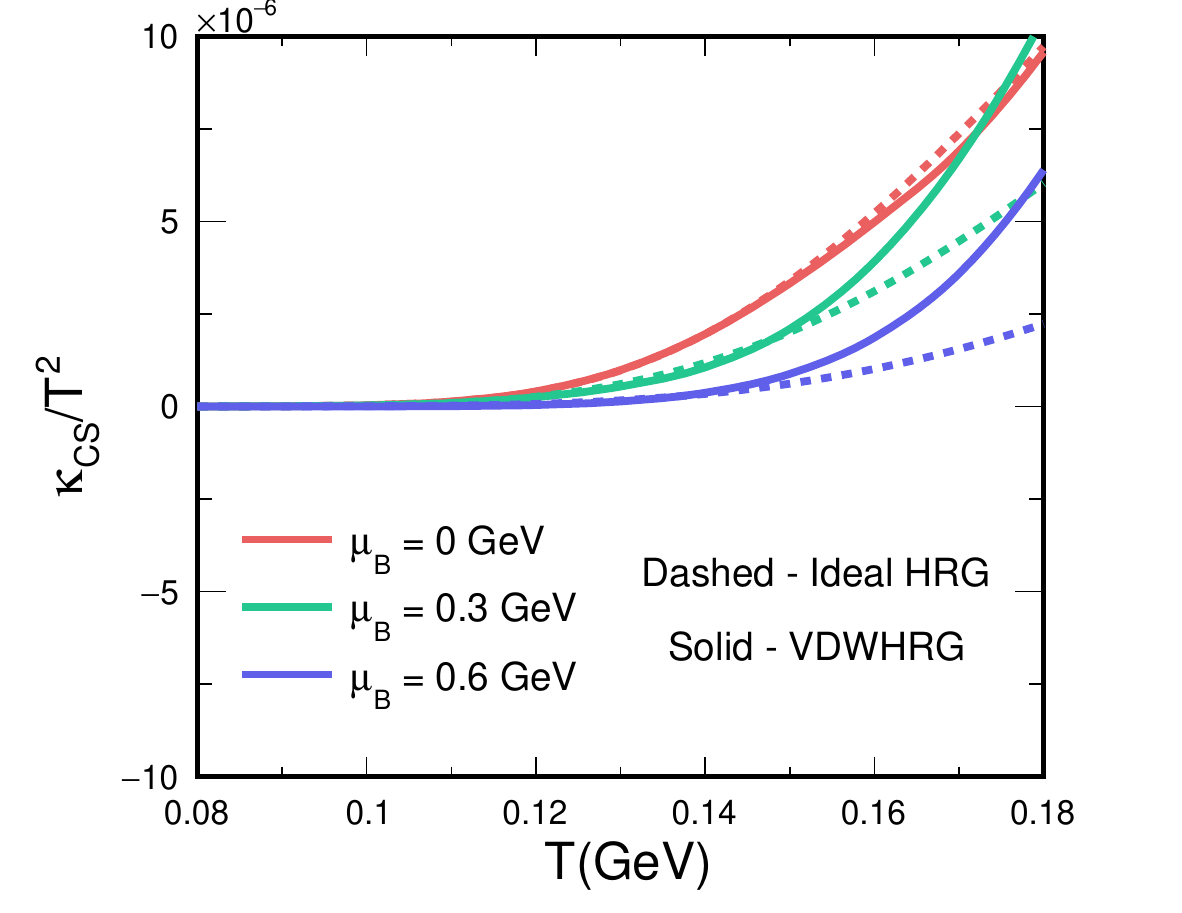}
      \caption{The off-diagonal terms of the diffusion matrix related to the baryon, electric charge, strangeness, and charm charge as a function of temperature for three different baryochemical potentials.}
      \label{fig-offdiag}
\end{figure*}
In Fig.~\ref{fig:EC}, we plot scaled electrical conductivity, $\kappa_{QQ}/T^{2} = \sigma_{el}/T$ as a function of temperature. We compare our results with many phenomenological estimations of $\sigma_{el}/T$.
In the hadronic sector, we observe a decrease in scaled electrical conductivity with temperature. Our results agree well with results from Kinetic theory~\cite{Greif:2016skc} and calculations from ChPT~\cite{Fernandez-Fraile:2005bew}. In the partonic phase, we plot the scaled electrical conductivity estimation from the Polyakov Nambu Jona-Lasinio model (PNJL)~\cite{Soloveva:2020hpr} and the dynamical quasiparticle model (DQPM)~\cite{Soloveva:2019xph}. Ref.~\cite{Cassing:2013iz} estimates conductivity in the hadronic to partonic phase utilizing the parton-hadron-string dynamics (PHSD) transport approach. Moreover, we compare our estimation with the lQCD data for $N_{f}=2+1$~\cite{Aarts:2014nba}.

In Fig.~\ref{fig-offdiag}, results for off-diagonal terms of diffusion matrix are depicted. From the order of the diffusion coefficients, the contribution of the off-diagonal components is as significant as that of the diagonal terms. The cross terms, for instance, $\kappa_{BQ}$, signifies that the gradient in charge chemical potential leads to diffusion in baryon quantum number because of the presence of charged baryon in the system. The opposite is also true for $\kappa_{QB}$ from the Onsager’s symmetry. In the cross terms, overall dependency on $T$ and $\mu_B$ are similar to that of diagonal terms. For example, the coefficients involved baryon quantum number ($\kappa_{BQ, ~BS, ~BC}$) increase with $\mu_B$. The value of $\kappa_{BS}$ is negative because strange particles have a negative strangeness number, and particle contributions dominate that of antiparticles. This negative contribution arising due to the negative strange quantum number affects other diffusion matrix coefficients, such as $\kappa_{QS}$. This effect becomes more pronounced with van der Waals interactions among the hadrons. However, in $\kappa_{QS}$, we get a competing effect from charged hadrons which has an interplay with strange charge resulting in a dip in $\kappa_{QS}$ value at $\mu_{B}$ = 0.6 GeV. Moreover, the negative diffusion coefficient signifies that the current of charges and the flow of particles are in opposite directions.

\begin{figure*}
    \includegraphics[width = 0.4\linewidth]{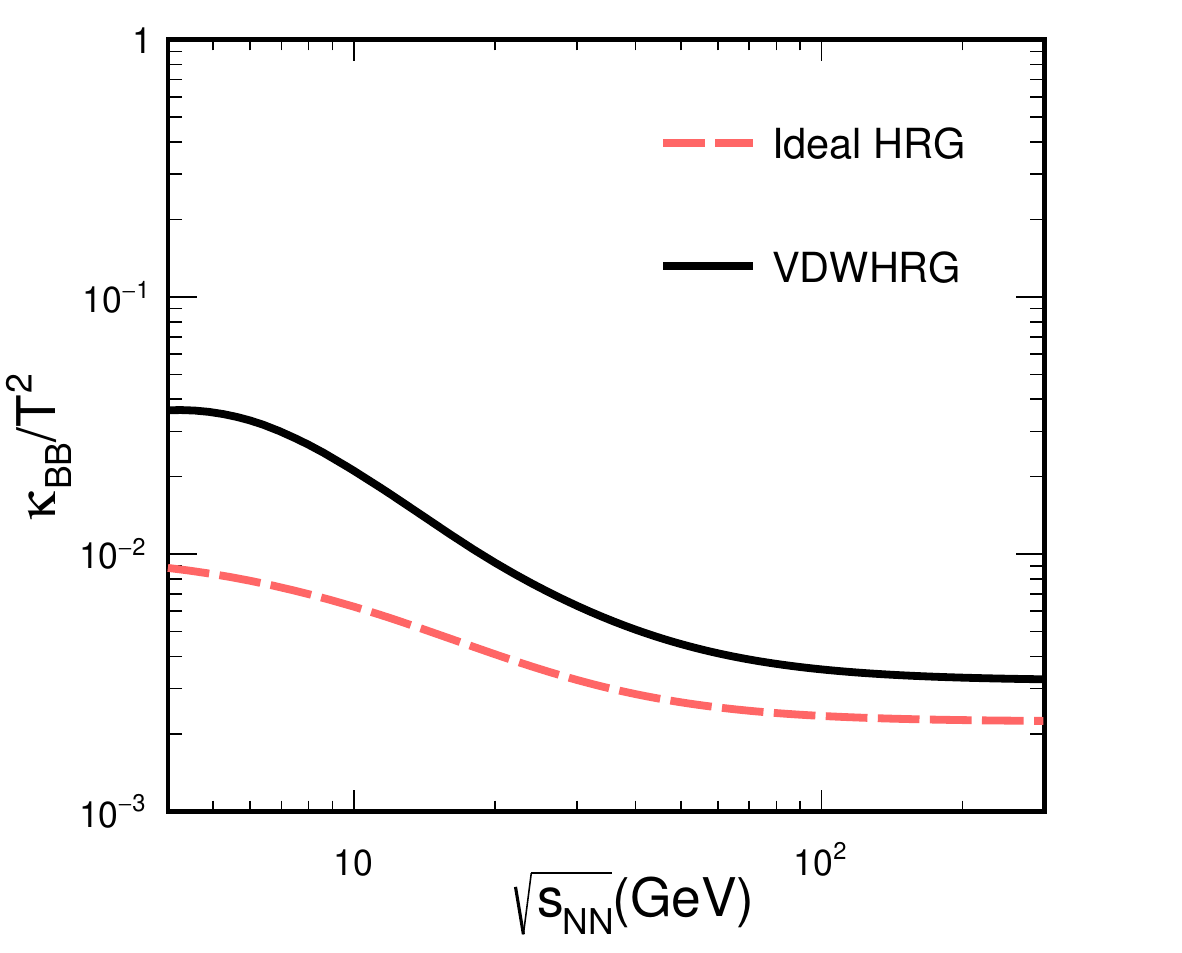}
    \includegraphics[width = 0.4\linewidth]{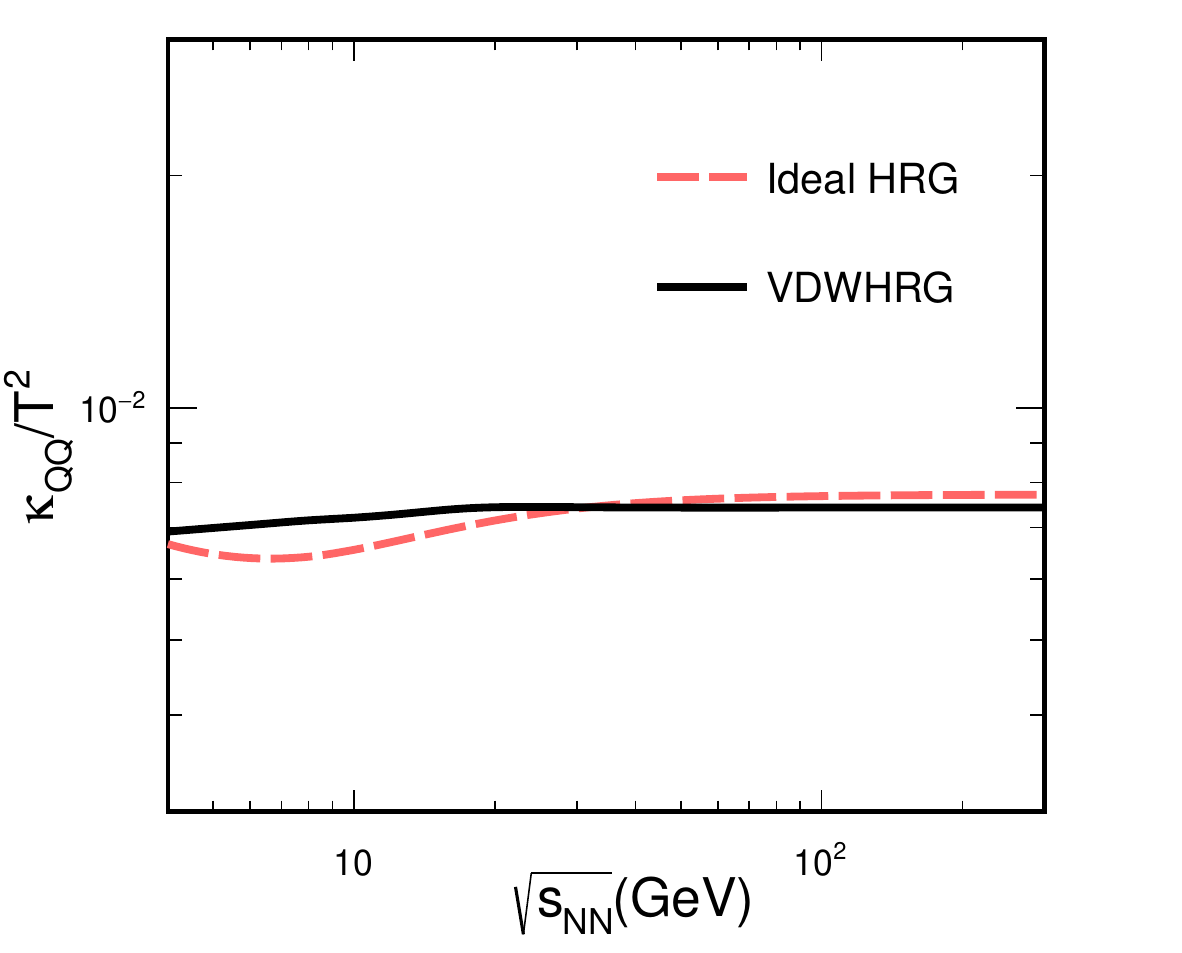}
    \includegraphics[width = 0.4\linewidth]{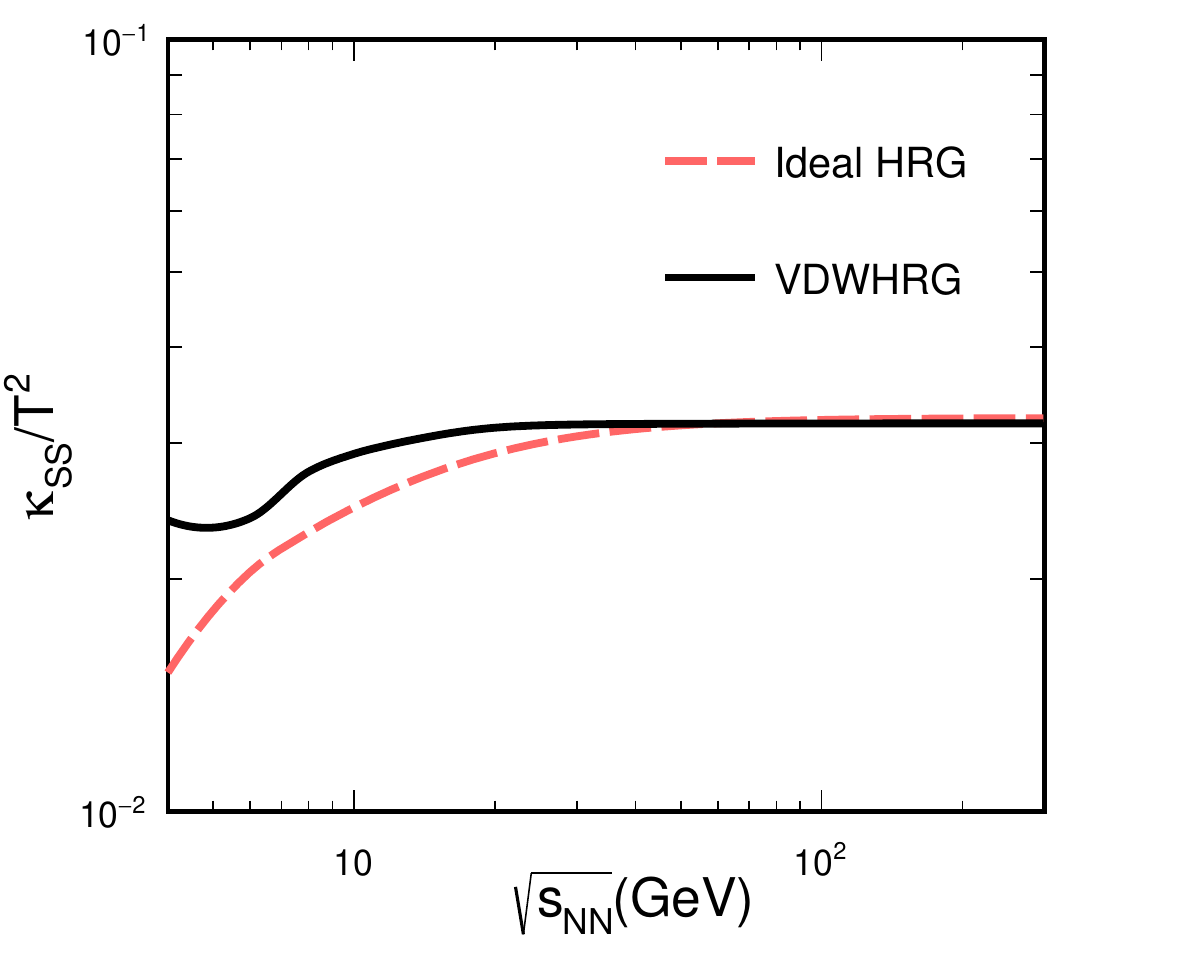}
    \includegraphics[width = 0.4\linewidth]{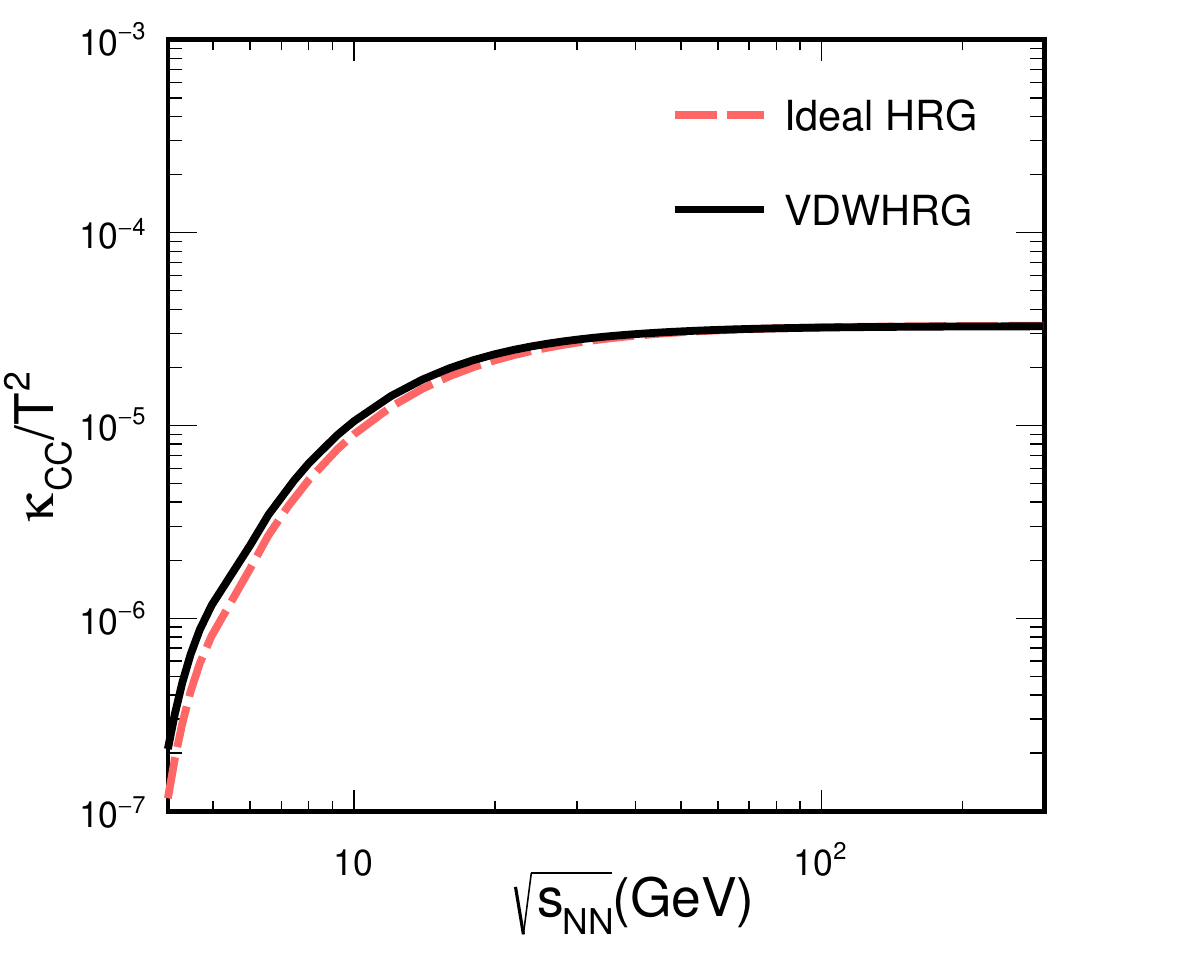}
    \caption{The diagonal terms of the diffusion matrix related to the baryon, electric charge, strangeness, and charm charge as a function of the center of mass energy.}
    \label{snn_figs}
\end{figure*}
The BES program at RHIC studies the proton number cumulants, which is a proxy for baryon number cumulants; kaon number cumulants, which is a proxy for strangeness cumulants; or pion, kaon, proton number cumulants, which are the proxy for electric charge cumulants~\cite{STAR:2022vlo, STAR:2020tga}. These are studied as functions of $\sqrt{s_{\rm NN}}$, which helps locate the QCD critical point. Thus, it is important to study the diffusion of conserved charges as a function of $\sqrt{s_{\rm NN}}$, which will give us an idea of how much the fluctuations are affected at a given center-of-mass energy. Fig.~\ref{snn_figs} shows the variation of the diffusion matrix's diagonal terms as a function of the center of mass energies.  On the top left panel, scaled $\kappa_{\rm BB}$ is plotted as a function of $\sqrt{s_{\rm NN}}$ for both ideal and van der Waals HRG cases. We observe that for both Ideal HRG and VDWHRG cases, the trend is decreasing throughout the $\sqrt{s_{\rm NN}}$ values and gets saturated after 100 GeV. The point to note here is that for low $\sqrt{s_{\rm NN}}$ values, the diffusion is maximum, which means the baryon number susceptibilities will be maximally affected. This means, that whatever proton number cumulants one gets from data is already contaminated from diffusion and will be a poor proxy to find the critical point. Similarly, we have plotted the scaled $\kappa_{\rm QQ}$ as a function of $\sqrt{s_{\rm NN}}$ on the top right panel, which remains nearly independent of the collision energy. In the bottom left panel, the scaled $\kappa_{\rm SS}$ is plotted as a function of $\sqrt{s_{\rm NN}}$. Here, the results from both models show saturation at very high center-of-mass energy. Moreover, $\kappa_{\rm CC}/T^{2}$ is plotted as a function of $\sqrt{s_{\rm NN}}$ in the bottom right panel. The trends for both ideal and VDWHRG are almost the same, increasing with $\sqrt{s_{\rm NN}}$ and getting saturated after around 20-30 GeV. The order of diffusion for charm is significantly lower than that of the other conserved charge diffusion, which is expected. From these, one can naively infer that to study the critical point through conserved charge susceptibilities; one should consider how the diffusion can affect the fluctuations till the final state. We have argued in our previous paper~\cite{Goswami:2023hdl}, a charm number fluctuation study through $D^{+}$, $D^{-}$ fluctuation is an interesting probe for phase transition. Here, it can be seen that due to the less diffusion of charm in the medium, charm fluctuation will remain relatively unaffected and thus will give a cleaner signal. Similarly, one can study the cross-terms of the diffusion matrix to gain important insights into how the diffusion can affect the cross-terms fluctuations.

 Here, it is essential to stress the applicability of RTA in the vicinity of a critical point. The correlation length ($\xi$) becomes very large at the critical point, and the Knudsen number, proportional to $\xi$, goes above unity. Therefore, the system's hydrodynamic (or continuum) description breaks down at a critical point, and RTA is not applicable there. For a detailed, quantitative description of high Knudsen numbers at critical points, see Refs.~\cite{Stephanov:2017ghc, Du:2021zqz}.

\section{Summary}

\label{summary}
In this paper, we deploy the van der Waals hadron resonance gas model to estimate the diagonal and off-diagonal diffusion matrix coefficient. In recent times, there have been significant advancements towards the thermalization of the charm sector. Following a similar notion, we keep the charmed hadrons on a similar footing as the lighter hadrons and study the diffusion matrix coefficient of the charmed charges as well as for baryon, strange, and electrical charge. We estimate these coefficients as a function of temperature for three different chemical potential values. We observe an interplay between relaxation time, total number density, and net charge density affects all the components of the diffusion matrix. For the diagonal components of baryon, strangeness, and electrical conserved charges, we observe a higher magnitude as compared to the $\kappa_{CC}/T^{2}$ values. This indicates a weak charm current in the hadronic medium. Moreover, we study the off-diagonal components of the diffusion matrix with temperature and at different chemical potentials. We observe that the off-diagonal diffusion matrix terms, such as $\kappa_{BQ}$ and $\kappa_{BS}$, are significant and show similar dependencies on $T$ and $\mu_B$ as the diagonal terms. We also observe a negative diffusion coefficient value for off-diagonal terms involving strange charges. This is because the strange particles have a negative strangeness charge, and with increasing chemical potential, there is an abundance of particles over anti-particles. Furthermore, we plot the diagonal components as a function of center-of-mass energy. We observe that at low $\sqrt{s_{NN}}$, the hadrons with conserved charges $B$, $Q$, and $S$ diffuse higher as compared to the charmed hadrons. Due to this, charm fluctuations created at the early stage of heavy-ion collisions will remain significant till the freeze-out. It hints toward a change in our approach to studying the location of the critical point, where net proton number cumulants are taken as a proxy for net baryon fluctuations. However, the higher diffusion of baryons would lead to the smeared signal of the critical point. This allows us to propose a new probe for the hunt of the critical point, i.e., the net charm number fluctuations, which is affected relatively less by diffusion and should be a better probe. However, it is worth noting here that given the possible location of the QCD critical point in the phase diagram, which
pertains to lower collision energies, one expects the charm production cross-section to be on the lower side on an event-by-event basis.
This will bring the issues of statistical fluctuation dominance in the game, while on the other side, as expected, critical points
will be associated with higher dynamical fluctuations. This will make the charm number fluctuation measurement more challenging while keeping it highly exciting.

\section*{Acknowledgments}
K.G. acknowledges the financial support from the Prime Minister's Research Fellowship (PMRF), Government of India. K.P. acknowledges the financial aid from UGC, Government of India. The authors gratefully acknowledge the DAE-DST, Government of India funding under the mega-science project “Indian participation in the ALICE experiment at CERN” bearing Project No. SR/MF/PS-02/2021-IITI(E-37123).
    
%


\begin{thebibliography}{}

\bibitem{Aoki:2006we}
Y.~Aoki, G.~Endrodi, Z.~Fodor, S.~D.~Katz and K.~K.~Szabo,
Nature \textbf{443}, 675 (2006).

\bibitem{Stephanov:1998dy}
M.~A.~Stephanov, K.~Rajagopal and E.~V.~Shuryak,
Phys. Rev. Lett. \textbf{81}, 4816 (1998).

\bibitem{STAR:2010mib}
M.~M.~Aggarwal \textit{et al.} [STAR],
Phys. Rev. Lett. \textbf{105}, 022302 (2010).

\bibitem{STAR:2013gus}
L.~Adamczyk \textit{et al.} [STAR],
Phys. Rev. Lett. \textbf{112}, 032302 (2014).

\bibitem{STAR:2014egu}
L.~Adamczyk \textit{et al.} [STAR],
Phys. Rev. Lett. \textbf{113}, 092301 (2014).

\bibitem{Asakawa:2000wh}
M.~Asakawa, U.~W.~Heinz and B.~Muller,
Phys. Rev. Lett. \textbf{85}, 2072 (2000).

\bibitem{Jeon:2000wg}
S.~Jeon and V.~Koch,
Phys. Rev. Lett. \textbf{85}, 2076 (2000).



\bibitem{HotQCD:2017qwq}
A.~Bazavov \textit{et al.} [HotQCD],
Phys. Rev. D \textbf{96}, 074510 (2017).


\bibitem{HotQCD:2012fhj}
A.~Bazavov \textit{et al.} [HotQCD],
Phys. Rev. D \textbf{86}, 034509 (2012).


\bibitem{Shuryak:2000pd}
E.~V.~Shuryak and M.~A.~Stephanov,
Phys. Rev. C \textbf{63}, 064903 (2001).

\bibitem{Fotakis:2019nbq}
J.~A.~Fotakis, M.~Greif, C.~Greiner, G.~S.~Denicol and H.~Niemi,
Phys. Rev. D \textbf{101}, 076007 (2020).

\bibitem{Greif:2017byw}
M.~Greif, J.~A.~Fotakis, G.~S.~Denicol and C.~Greiner,
Phys. Rev. Lett. \textbf{120}, 242301 (2018).

\bibitem{Fotakis:2021diq}
J.~A.~Fotakis, O.~Soloveva, C.~Greiner, O.~Kaczmarek and E.~Bratkovskaya,
Phys. Rev. D \textbf{104}, 034014 (2021).

\bibitem{Das:2021bkz}
A.~Das, H.~Mishra and R.~K.~Mohapatra,
Phys. Rev. D \textbf{106}, 014013 (2022).

\bibitem{Andronic:2005yp}
A.~Andronic, P.~Braun-Munzinger and J.~Stachel,
Nucl. Phys. A \textbf{772}, 167 (2006).

\bibitem{Borsanyi:2012}
S. Borsanyi, Z. Fodor, S. D. Katz, S. Krieg, C. Ratti, and K.
Szabo,
J. High Energy Phys. \textbf{01}, 138 (2012).



\bibitem{Bellwied:2015lba}
R.~Bellwied, S.~Borsanyi, Z.~Fodor, S.~D.~Katz, A.~Pasztor, C.~Ratti and K.~K.~Szabo,
Phys. Rev. D \textbf{92}, 114505 (2015).

\bibitem{Bellwied:2013cta}
R.~Bellwied, S.~Borsanyi, Z.~Fodor, S.~D.~Katz and C.~Ratti,
Phys. Rev. Lett. \textbf{111}, 202302 (2013).

\bibitem{Bazavov:2013dta}
A.~Bazavov, H.~T.~Ding, P.~Hegde, O.~Kaczmarek, F.~Karsch, E.~Laermann, Y.~Maezawa, S.~Mukherjee, H.~Ohno and P.~Petreczky, \textit{et al.}
Phys. Rev. Lett. \textbf{111}, 082301 (2013).

\bibitem{Braun-Munzinger:1999hun}
P.~Braun-Munzinger, I.~Heppe and J.~Stachel,
Phys. Lett. B \textbf{465}, 15 (1999).

\bibitem{Bhattacharyya:2013oya}
A.~Bhattacharyya, S.~Das, S.~K.~Ghosh, R.~Ray and S.~Samanta,
Phys. Rev. C \textbf{90}, 034909 (2014).

\bibitem{Vovchenko:2016rkn}
V.~Vovchenko, M.~I.~Gorenstein and H.~Stoecker,
Phys. Rev. Lett. \textbf{118}, 182301 (2017).

\bibitem{Samanta:2017yhh}
S.~Samanta and B.~Mohanty,
Phys. Rev. C \textbf{97}, 015201 (2018).

\bibitem{Pradhan:2022gbm}
K.~K.~Pradhan, D.~Sahu, R.~Scaria and R.~Sahoo,
Phys. Rev. C \textbf{107}, 014910 (2023).

\bibitem{Sahoo:2023vkw}
B.~Sahoo, K.~K.~Pradhan, D.~Sahu and R.~Sahoo,
Phys. Rev. D \textbf{108}, 074028 (2023).

\bibitem{Pradhan:2023rvf}
K.~K.~Pradhan, B.~Sahoo, D.~Sahu and R.~Sahoo,
Eur. Phys. J. C \textbf{84}, 936 (2024).

\bibitem{Bazavov:2014yba}
A.~Bazavov, H.~T.~Ding, P.~Hegde, O.~Kaczmarek, F.~Karsch, E.~Laermann, Y.~Maezawa, S.~Mukherjee, H.~Ohno and P.~Petreczky, \textit{et al.}
Phys. Lett. B \textbf{737}, 210 (2014).

\bibitem{Goswami:2023hdl}
K.~Goswami, K.~K.~Pradhan, D.~Sahu and R.~Sahoo,
Phys. Rev. D \textbf{108}, 074011 (2023).

\bibitem{ALICE:2017quq}
S.~Acharya \textit{et al.} [ALICE],
Phys. Rev. Lett. \textbf{119}, 242301 (2017).

\bibitem{CMS:2017vhp}
A.~M.~Sirunyan \textit{et al.} [CMS],
Phys. Rev. Lett. \textbf{120}, 202301 (2018).

\bibitem{ATLAS:2017xqp}
M.~Aaboud \textit{et al.} [ATLAS],
Phys. Lett. B \textbf{780}, 578 (2018)

\bibitem{ALICE:2020pvw}
S.~Acharya \textit{et al.} [ALICE],
JHEP \textbf{10}, 141 (2020)

\bibitem{He:2021zej}
M.~He, B.~Wu and R.~Rapp,
Phys. Rev. Lett. \textbf{128}, 162301 (2022).

\bibitem{Wu:2023djn}
B.~Wu, Z.~Tang, M.~He and R.~Rapp,
Phys. Rev. C \textbf{109}, 014906 (2024)

\bibitem{vanHees:2004gq}
H.~van Hees and R.~Rapp,
Phys. Rev. C \textbf{71}, 034907 (2005).

\bibitem{Goswami:2022szb}
K.~Goswami, D.~Sahu and R.~Sahoo,
Phys. Rev. D \textbf{107}, 014003 (2023).


\bibitem{HotQCD:2014kol}
A.~Bazavov \textit{et al.} [HotQCD],
Phys. Rev. D \textbf{90}, 094503 (2014).


\bibitem{Borsanyi:2013bia}
S.~Borsanyi, Z.~Fodor, C.~Hoelbling, S.~D.~Katz, S.~Krieg and K.~K.~Szabo,
Phys. Lett. B \textbf{730}, 99 (2014).

\bibitem{Sarkar:2018mbk}
N.~Sarkar and P.~Ghosh,
Phys. Rev. C \textbf{98}, 014907 (2018).

\bibitem{Pal:2021qav}
S.~Pal, G.~Kadam and A.~Bhattacharyya,
Nucl. Phys. A \textbf{1023}, 122464 (2022).

\bibitem{Dashen:1969ep}
R.~Dashen, S.~K.~Ma and H.~J.~Bernstein,
Phys. Rev. \textbf{187}, 345 (1969).

\bibitem{Andronic:2012ut}
A.~Andronic, P.~Braun-Munzinger, J.~Stachel and M.~Winn,
Phys. Lett. B \textbf{718}, 80 (2012).

\bibitem{Vovchenko:2015vxa}
V.~Vovchenko, D.~V.~Anchishkin and M.~I.~Gorenstein, Phys. Rev. C \textbf{91}, 064314 (2015).

\bibitem{Vovchenko:2015pya}
V.~Vovchenko, D.~V.~Anchishkin, M.~I.~Gorenstein and R.~V.~Poberezhnyuk, Phys. Rev. C \textbf{92}, 054901 (2015).

\bibitem{Poberezhnyuk:2019pxs}
R.~Poberezhnyuk, V.~Vovchenko, A.~Motornenko, M.~I.~Gorenstein and H.~Stoecker,
Phys. Rev. C \textbf{100}, 054904 (2019).

\bibitem{Becattini:2000jw}
F.~Becattini, J.~Cleymans, A.~Keranen, E.~Suhonen and K.~Redlich,
Phys. Rev. C \textbf{64}, 024901 (2001).

\bibitem{Cleymans:1998yb}
J.~Cleymans, H.~Oeschler and K.~Redlich,
Phys. Rev. C \textbf{59}, 1663 (1999)


\bibitem{Begun:2018qkw}
V.~V.~Begun, V.~Vovchenko, M.~I.~Gorenstein and H.~Stoecker,
Phys. Rev. C \textbf{98}, 054909 (2018)


\bibitem{Albright:2015fpa}
M.~Albright and J.~I.~Kapusta,
Phys. Rev. C \textbf{93}, 014903 (2016).

\bibitem{Chakraborty:2010fr}
P.~Chakraborty and J.~I.~Kapusta,
Phys. Rev. C \textbf{83} (2011), 014906

\bibitem{Dey:2019axu}
J.~Dey, S.~Satapathy, P.~Murmu and S.~Ghosh,
Pramana \textbf{95} (2021) 125


\bibitem{Chapman_book}
S. Chapman and T. G. Cowling, The Mathematical
Theory of Non-Uniform Gases: An Account of the Kinetic
Theory of Viscosity, Thermal Conduction and Diffusion in
Gases (Cambridge University Press, Cambridge, England,
1970).

\bibitem{Kadam:2015xsa}
G.~P.~Kadam and H.~Mishra,
Phys. Rev. C \textbf{92}, 035203 (2015).


\bibitem{Cleymans:2005xv}
J.~Cleymans, H.~Oeschler, K.~Redlich and S.~Wheaton,
Phys. Rev. C \textbf{73}, 034905 (2006).

\bibitem{Behera:2022nfn}
S.~P.~Behera and D.~K.~Mishra,
Nucl. Phys. A \textbf{1024}, 122475 (2022).

\bibitem{Tiwari:2011km}
S.~K.~Tiwari, P.~K.~Srivastava and C.~P.~Singh,
Phys. Rev. C \textbf{85}, 014908 (2012).

\bibitem{Greif:2016skc}
M.~Greif, C.~Greiner and G.~S.~Denicol,
Phys. Rev. D \textbf{93}, 096012 (2016)
[erratum: Phys. Rev. D \textbf{96}, 059902 (2017)].

\bibitem{Fernandez-Fraile:2005bew}
D.~Fernandez-Fraile and A.~Gomez Nicola,
Phys. Rev. D \textbf{73}, 045025 (2006).

\bibitem{Soloveva:2020hpr}
O.~Soloveva, D.~Fuseau, J.~Aichelin and E.~Bratkovskaya,
Phys. Rev. C \textbf{103}, 054901 (2021)

\bibitem{Soloveva:2019xph}
O.~Soloveva, P.~Moreau and E.~Bratkovskaya,
Phys. Rev. C \textbf{101}, 045203 (2020)

\bibitem{Cassing:2013iz}
W.~Cassing, O.~Linnyk, T.~Steinert and V.~Ozvenchuk,
Phys. Rev. Lett. \textbf{110}, 182301 (2013).

\bibitem{Aarts:2014nba}
G.~Aarts, C.~Allton, A.~Amato, P.~Giudice, S.~Hands and J.~I.~Skullerud,
JHEP \textbf{02}, 186 (2015)

\bibitem{STAR:2020tga}
J.~Adam \textit{et al.} [STAR],
Phys. Rev. Lett. \textbf{126}, 092301 (2021).

\bibitem{STAR:2022vlo}
B.~Aboona \textit{et al.} [STAR],
Phys. Rev. Lett. \textbf{130}, 082301 (2023).

\bibitem{Stephanov:2017ghc}
M.~Stephanov and Y.~Yin,
Phys. Rev. D \textbf{98}, 036006 (2018).

\bibitem{Du:2021zqz}
L.~Du, X.~An and U.~Heinz,
Phys. Rev. C \textbf{104}, 064904 (2021).




\end{thebibliography}
\end{document}